\documentclass[sigconf,nonacm]{acmart}

\setcopyright{none}

\copyrightyear{}
\acmYear{}
\acmDOI{}

\acmConference[]{ }{ }{ }

\usepackage{graphicx}
\usepackage{listings}
\usepackage{xcolor}
\usepackage{multicol}
\usepackage{tikz}
\usetikzlibrary{arrows}

\newcommand{\CommentColor}{\color{white!60!black}}
\newcommand{\StringColor}{\color{green!70!black}}
\newcommand{\UserKeywordsColor}{\color{red!60!black}}
\newcommand{\KeywordsColor}{\color{violet}}

\newcommand{\sparqlkw}[1]{{\UserKeywordsColor\textbf{\texttt{#1}}}}
\newcommand{\ourkw}[1]{{\KeywordsColor\textbf{\texttt{#1}}}}

\lstdefinelanguage{SPARQL}
{
	basicstyle=\footnotesize\ttfamily,
	keywordstyle=\KeywordsColor\textbf,
	keywordstyle=[2]\UserKeywordsColor\textbf,,
	tabsize=2,
	morekeywords={LET, DO, WHILE, PRINT, QVALUES, FIXPOINT, TIMES, RETURN},
	morekeywords=[2]{SELECT, UNION, MINUS, WHERE, ASK, BIND, COALESCE, OPTIONAL, GROUP, BY, AS, COUNT, ORDER, LIMIT, SUM, DISTINCT, FILTER, NOT, EXISTS, MINUS, VALUES, AND, DESC, WITH, RECURSIVE},
	commentstyle=\CommentColor,
	stringstyle=\StringColor,
	sensitive=true,
	morecomment=[l]{\#},
	morecomment=[s]{/*}{*/},
	morestring=[b]",
	showstringspaces=false,
	aboveskip=3pt, 
	belowskip=3pt,
	mathescape=true,
	extendedchars=true,
    literate={á}{{\'a}}1 {ã}{{\~a}}1 {é}{{\'e}}1 {í}{{\'i}}1 {ó}{{\'o}}1 {ú}{{\'u}}1 {ñ}{{\~n}}1,
    numbers=left,
  	firstnumber=1,
  	numberfirstline=true,
  	xleftmargin=20pt
}

\newcommand{\querylytics}{queralytics}

\newcommand{\paralist}[1]{\smallskip\noindent\textbf{#1}:}

\newcommand{\Pro}{P}
\newcommand{\qvalues}{\ourkw{QVALUES}}
\newcommand{\values}{\text{val}}

\newcommand{\COM}{\operatorname{COM}}
\newcommand{\AGG}{\operatorname{AGG}}
\newcommand{\lms}{\{\!\!\{}
\newcommand{\rms}{\}\!\!\}}
\newcommand{\vx}{\textbf{x}}

\usepackage{subcaption}

\begin{document}

\title{Recursive SPARQL for Graph Analytics}

\author{Aidan Hogan}
\email{aidhog@gmail.com}
\affiliation{%
  \institution{Universidad de Chile \& IMFD}
}

\author{Juan Reutter}
\email{jreutter@ing.puc.cl}
\affiliation{%
  \institution{PUC Chile \& IMFD}
}

\author{Adri\'an Soto}
\email{assoto@uc.cl}
\affiliation{%
  \institution{PUC Chile \& IMFD}
}

\renewcommand{\shortauthors}{Hogan, et al.}

\begin{abstract}
Work on knowledge graphs and graph-based data management often focus either on declarative graph query languages or on frameworks for graph analytics, where there has been little work in trying to combine both approaches. However, many real-world tasks conceptually involve combinations of these approaches: a graph query can be used to select the appropriate data, which is then enriched with analytics, and then possibly filtered or combined again with other data by means of a query language. In this paper we propose a declarative language that is well suited to perform graph querying and analytical tasks. We do this by proposing a minimalistic extension of SPARQL to allow for expressing analytical tasks; in particular, we propose to extend SPARQL with recursive features, and provide a formal syntax and semantics for our language. We show that this language can express key analytical tasks on graphs (in fact, it is Turing complete), offering a more declarative alternative to existing frameworks and languages. We show how procedures in our language can be implemented over an off-the-shelf SPARQL engine with a specialised client that allows parallelisation and batch-based processing when memory is limited. Results show that with such an implementation, procedures for popular analytics currently run in seconds or minutes for selective sub-graphs (our target use-case) but struggle at larger scales.
\end{abstract}

\newcommand{\ah}[1]{{\color{blue}\textsc{ah:} #1}}
\newcommand{\jr}[1]{{\color{red}\textsc{jr:} #1}}

%


\maketitle

\section{Introduction}

Recent years have seen a surge in interest in graph data management, learning and analytics spanning various academic communities. Much of this work has been conducted under the title of ``knowledge graphs''~\cite{GoogleKG}, centred on the composition and extraction of knowledge from graph-structured data at large-scale, drawing together techniques from communities such as Databases, Graph Theory, Machine Learning, the Semantic Web, and more besides~\cite{BonattiDPP18}. A variety of major commercial websites are now using proprietary knowledge graphs to support various applications~\cite{AirBnBKG,AmazonKG,BingKG,eBayKG,GoogleKG,LinkedInKG}. Non-proprietary knowledge graphs like Wikidata~\cite{VrandecicK14} -- published on the Web using Semantic Web standards -- have been widely adopted for numerous applications. Wikidata's SPARQL query service now receives millions of queries per day~\cite{MalyshevKGGB18}. 

However, while works on knowledge graphs are currently being pursued by various communities, more work is needed to combine complementary techniques from different areas~\cite{BonattiDPP18}. As a prominent example, while a variety of query languages have been proposed for graphs~\cite{sparql11,Rodriguez15,FrancisGGLLMPRS18,AnglesABBFGLPPS18,AnglesABHRV17}, and a variety of frameworks have been proposed for graph analytics~\cite{MalewiczABDHLC10,XinGFS13,signalcollect}, there are few works that aim to combine both querying and analytics for graphs: while some analytical frameworks support lightweight query features~\cite{XinGFS13,Rodriguez15}, and some query languages support lightweight analytical features~\cite{sparql11,FrancisGGLLMPRS18}, these solutions are limited to specific types of queries, or specific analytics, or require imperative ``glue'' code. We argue that a more general declarative alternative is needed.

Take, for example, the following seemingly simple task, which we wish to apply over Wikidata: \textit{find stations from which one can still reach Palermo metro station in Buenos Aires if Line C is closed}. Although standard graph query languages -- such as SPARQL~\cite{sparql11}, Cypher~\cite{FrancisGGLLMPRS18}, G-CORE~\cite{AnglesABBFGLPPS18}, etc. -- support path expressions that capture reachability, they cannot express conditions on the nodes through which such paths pass, as is required by this task (i.e., that they are not on Line C). Consider a more complex example that again, in principle, can be answered over Wikidata: \textit{find the top author of scientific articles about the Zika virus according to their $p$-index within the topic}. The $p$-index of authors is calculated by computing the PageRank of papers in the citation network, and then summing the scores of the papers for each respective author~\cite{Senanayake15}. One way this could currently be achieved is to: (1) perform a SPARQL query to extract the citation graph of scientific articles about the Zika virus; (2) load the graph into an external tool to compute PageRank scores; (3) perform another query to extract the (bipartite) authorship graph for the articles; (4) load the authorship graph into the external tool to join authors with papers, aggregate the $p$-index score per author, sort by score, and output the top result. Here the user must ship data back and forth between different tools to solve the task. Another strategy might be to load the Wikidata dump into a graph-analytics framework, writing code to extract the required graphs, analyse them, and aggregate the results; in this case, we lose the convenience of a declarative query language and database optimisations for extracting the relevant data, performing joins and aggregations, etc., as the task requires.

In this paper, we instead propose a general, (mostly) declarative language that supports \textit{graph queralytics}: tasks that combine querying and analytics on graphs, allowing to interleave both arbitrarily. We coin the term ``\textit{queralytics}'' to highlight that these tasks raise new challenges and are not well-supported by existing languages and tools that focus only on querying or analytics. Rather than extending a graph query language with support for specific, built-in analytics, we rather propose to extend a graph query language to be able to express any form of (computable) analytical task of interest to the user: namely we add recursion to the query language. Specifically, we explore the addition of recursive features to the SPARQL query language, proposing a concrete syntax and semantics for our language, showing examples of how it can combine querying and analytics for graphs. We call our language the \textit{SPARQL Protocol and RDF Query \& Analtyics Language} (\textit{SPARQAL}). We study the expressive power of SPARQAL with similar proposals found in the literature~\cite{ReutterSV15,CorbyFG17,Urzua019}. We then discuss the implementation of our language on top of a SPARQL query engine, introducing evaluation strategies that aim to find trade-offs between scalability and performance. We present experiments to compare our proposed strategies on real-world datasets, for which we devise a set of benchmark queralytics over Wikidata. Our results provide insights into the scale and performance with which an existing SPARQL engine can perform standard graph analytics, showing that for queralytics wherein a selective sub-graph is extracted for analysis, interactive performance is feasible; on the other hand, the current implementation struggles for an analytical benchmark on a larger-scale graph.

\begin{example}\label{ex:metro}
To illustrate our proposal, we provide a queralytic in our language for the first example seen in the introduction (the Zika/$p$-index task will be seen later). Namely, suppose that there is a concert close to Palermo metro station in Buenos Aires; however, Line C of the metro is closed due to a strike. We would like to know from which metro stations we can still reach Palermo. We can express this queralytic in our SPARQL-based language as follows:

\begin{lstlisting}
LET reachable = ( 
  SELECT ?s WHERE { 
    wd:Q3296629 wdt:P197 ?s .
    MINUS { ?s wdt:P81 wd:Q1157050 }
  } 
);
DO (
  LET adjacent = (
    SELECT (?adj AS ?s) WHERE { 
      ?s wdt:P197 ?adj . 
      MINUS { ?adj wdt:P81 wd:Q1157050 } 
      QVALUES(reachable) 
    }
  );
  LET reachable = ( 
    SELECT DISTINCT ?s WHERE { 
      { QVALUES(adjacent) } 
      UNION 
      { QVALUES(reachable) } 
    }
  );
) WHILE( FIXPOINT(reachable) );
RETURN(reachable);
\end{lstlisting}

Here we work with the Wikidata dataset, where two adjacent stations are given by the property \texttt{wdt:P197} and the metro line by \texttt{wdt:P81}; the entities \texttt{wd:Q3296629} and \texttt{wd:Q1157050} refer to Palermo metro station and Line C, respectively. From lines 1 to 6, we first define a \textit{solution variable} called \texttt{reachable} whose value is the result of computing all stations directly adjacent to Palermo that are not on Line C. From lines 7 to 22 we have a loop that executes two instructions: the first, starting at line 8, computes all stations directly adjacent to the current reachable stations not on Line C; the second, starting at line 15, adds the new adjacent stations to the list of known reachable stations with a union. 
The loop is finished when the set of solutions assigned to the variable \texttt{reachable} does not change from one iteration to another (a fixpoint is thus reached). Finally, on line 23, we return the reachable stations. \qed
\end{example}

\section{Related Work}
In terms of related works, we first discuss frameworks and languages for applying graph analytics. We then discuss prior proposals for combining graph querying and graph analytics. We then introduce works on extending graph query languages with recursion. We end by highlighting the novelty of this work.

\paragraph{Frameworks for Graph Analytics} Given the growing need to perform graph analytics at large-scale -- involving the Web, social networks, etc. -- various frameworks have been proposed for such settings, including GraphStep~\cite{DeLorimierKMRERUKD06}, Pregel~\cite{MalewiczABDHLC10}, HipG~\cite{KrepskaKFB11}, PowerGraph~\cite{GonzalezLGBG12}, GraphX~\cite{XinGFS13}, Giraph~\cite{ChingEKLM15},  Signal/Collect~\cite{signalcollect}, and more besides. All such frameworks operate on a computational model -- sometimes called the systolic model~\cite{LowGKBGH14}, Gather/Apply/Scatter (GAS) model~\cite{GonzalezLGBG12}, graph-parallel framework~\cite{XinGFS13}, etc. -- that involves each node in a graph recursively computing its state based on data available for its neighbouring nodes according to a given function. Although such frameworks allow for large-scale graph analytics to be applied in a distributed setting, implementing queries on such frameworks, selecting custom sub-graphs to be analysed, etc., is not straightforward. Similar computational models are used in the case of graph neural networks~\cite{ScarselliGTHM09,abs-1901-00596}, which have been shown to be as discriminative as the (incomplete) Weisfeiler--Lehman (WL) graph isomorphism test~\cite{XuHLJ19}: in other words, by basing computation only on local information in each node's neighbourhood, there are certain pairs of non-isomorphic graphs that will return ``isomorphic results'' for any algorithm implemented in the framework. 

\paragraph{Graph Queries and Analytics}

Our work aims to combine graph queries and analytics, focusing on RDF graphs. One such proposal along these lines is Trinity.RDF~\cite{ZengYWSW13}, which stores RDF in a native graph format where nodes store inward and outward adjacency lists, allowing to traverse from a node to its neighbours without the need for index lookup; the system is then implemented in a distributed in-memory index, with query processing and optimisation components provided for basic graph patterns. Although the authors discuss how Trinity.RDF's storage scheme can also be useful for graph algorithms based on random walks, reachability, etc., experiments focus on SPARQL query evaluation from standard benchmarks~\cite{ZengYWSW13}. Later work used the same infrastructure in a system called Trinity~\cite{ShaoWL13} to implement and perform experiments with respect to PageRank and Breadth-First Search, this time rather focusing on graph analytics without performing queries. Though such an infrastructure could be adapted to apply graph queralytics at scale, the authors do not discuss the combination of queries and analytics, nor do they propose languages along these lines.

Most modern graph query languages directly support some built-in analytical features. SPARQL 1.1~\cite{sparql11} introduced \textit{property paths}~\cite{KostylevR0V15} that allow for specifying regular expressions on paths; these can then be used in the context of a SPARQL query to find pairs of nodes connected by some path matching the regular expression. The Cypher query language for property graphs~\cite{FrancisGGLLMPRS18} (used by the Neo4j graph database~\cite{Miller13}) also allows for querying on paths; though limited in terms of the regular expressions it allows on paths when compared to SPARQL 1.1, it offers features that SPARQL 1.1 does not, including shortest paths, returning paths, etc. The G-CORE query language~\cite{AnglesABBFGLPPS18} also supports features relating to paths, allowing to store and label paths, find weighted shortest paths, and more besides. In general, however, graph query languages tend to only support analytics relating to path finding and reachability~\cite{AnglesABHRV17}.

The Gremlin language~\cite{Rodriguez15} is more imperative in style than the aforementioned query languages, allowing to express analytical tasks through graph traversals. Per the Trinity.RDF system~\cite{ZengYWSW13}, graph traversals, when combined with variables, can be used to express and evaluate, for example, basic graph patterns~\cite{AnglesABBFGLPPS18}. Gremlin~\cite{Rodriguez15} also supports some declarative query operators, such as union, projection, negation, path expressions, and so forth, along with recursion, which allows to capture general analytical tasks; in fact, the Gremlin language is Turing complete~\cite{Rodriguez15}. 


In the context of SQL, languages such as Shark~\cite{XinRZFSS13} have been proposed that allow SQL queries to be embedded and executed in the context of distributed frameworks (in this case Spark~\cite{spark}) within which analytics can also be imperatively coded. Aside from embedding SQL into imperative languages, a number of languages have recently been proposed to combine relational algebra with linear algebra -- including LARA~\cite{HutchisonHS17} and MATLANG~\cite{BrijderGBW18} -- based on the observation that although relational algebra is often used for declarative querying, and linear algebra for learning and analytics, many operations in relational algebra can be simulated with linear algebra, and vice-versa, where it is thus of interest to understand the expressive power of both and how they complement each other~\cite{Geerts19}.

\paragraph{Recursive Graph Queries} Previous works have looked at adding recursive features to graph query languages. As aforementioned, most query languages support recursively matching path expressions in a graph; however, per Example~\ref{ex:metro}, more powerful forms of recursion are needed in the context of graph query languages to support the general class of analytics that we target here.\footnote{Though more complex forms of ``navigational patterns'' have been proposed in the literature, they are mostly limited to path-finding and reachability~\cite{AnglesABHRV17}.}

A number of authors have proposed more general recursion for graph query languages. Reutter et al.~\cite{ReutterSV15} propose to extend SPARQL with recursion based on \texttt{CONSTRUCT} queries; in particular, noting that \texttt{CONSTRUCT} transforms one RDF graph to another, they propose a syntax for recursively applying a \texttt{CONSTRUCT} template to the input graph up to a fixpoint, where a query can then be executed on the resulting fixpoint graph; they further propose a \textit{linear recursive} fragment of their language, which assumes that in each iteration only the data from the original graph and the previous iteration are required, reducing the complexity of evaluation. In later work, Corby et al.~\cite{CorbyFG17} proposed the LDScript language, which supports the definition of functions using SPARQL expressions; local variables that can store individual values, lists or the results of queries; and iteration over lists of values using loops, as well as recursive function calls. Recently Urzua and Gutierrez~\cite{Urzua019} proposed an extension of the G-CORE language to support linear recursion, and show how the resulting language can be used in principle to express various graph algorithms, such as a topological sort, which cannot be expressed in G-CORE without recursion.

\paragraph{Novelty} Unlike graph analytics frameworks, we propose a language for combining queries and analytics on graphs. Compared with Gremlin, our language is more declarative, based on an extension of an existing query language (SPARQL) to allow for expressing and combining graph analytics and queries. The closest proposals to ours are those that extend graph query languages with recursive features~\cite{ReutterSV15,CorbyFG17,Urzua019}. In comparison with the proposal of Reutter et al.~\cite{ReutterSV15} and Urzua and Gutierrez~\cite{Urzua019}, we allow recursion over \texttt{SELECT} queries, which adds flexibility by not requiring to maintain intermediate results as (RDF) graphs: for example, with \texttt{SELECT} we can maintain a table of four columns/variables representing a weighted RDF graph, where the first three columns denote an RDF graph and the fourth column denotes weights on individual triples; in the case of \texttt{CONSTRUCT}, we would rather require some form of reification to capture weighted triples. Furthermore, while we support fixpoint recursion, we also support other forms of recursion; in particular, we allow for terminating a loop based on a boolean condition (an \texttt{ASK} query), which offers greater flexibility for defining termination conditions in cases where, for example, an analytics task is infinitary and/or requires approximation in practice (e.g., PageRank). In comparison with LDScript~\cite{CorbyFG17} -- which also supports recursion on \texttt{SELECT} queries -- our focus is rather on supporting graph analytics with such a language, supporting features, such as fixpoint, that are useful in this setting.


\section{Language}

Recursion stands out in the literature as a key feature for supporting graph analytics. Our proposal -- called SPARQAL -- extends SPARQL (1.1) with recursion by allowing to iteratively evaluate queries (optionally) joined with solution sequences of prior queries until some condition is met. In order to support this form of iteration, we need two key operators. First, we extend SPARQL with \textit{solution variables} to which the results of a SELECT query can be assigned, and which can then be used within other queries to join solutions. Second, we extend SPARQL with \textit{\texttt{do-while} loops} to support iteratively repeating a sequence of SPARQL queries until some termination condition is met; this condition may satisfy a fixed number of iterations, a boolean ASK query, or a fixpoint on a solution variable (terminating when the set of solutions do not change).

We refer back to Example~\ref{ex:metro}, which illustrates how our language can be used to address a relatively simple queralytic task. We now present the syntax of our language, and thereafter proceed to define the formal semantics. We finish the section with a second, more involved example for computing the $p$-index of authors in an area.

\paralist{Preliminaries} To formally define our language and give our examples we assume familiarity with SPARQL and basic notions of graph analytics algorithms. We use the standard syntax and semantics of SPARQL in terms of mappings~\cite{sparql11}. We recall the notion of a \textit{solution sequence}, which is the result of a SPARQL query evaluated on a graph (or dataset), listing the ways in which the query matches the data. There may be zero, one or multiple solutions to a query.

\subsection{Syntax}
SPARQAL aims to be a minimalistic extension of the SPARQL language that allows to express queralytic tasks. Specifically, a task is defined as a \textit{procedure}, which is a sequence of \textit{statements}. A statement can be an \textit{assignment}, \textit{loop} or \textit{return} statement, as follows.

\paralist{Assignment} Assigns the solution sequence of a query to a solution variable. The syntax of an assignment statement is: 
\begin{center}
\texttt{\ourkw{LET} var = (Q);}
\end{center}
\noindent where \texttt{var} is a variable name and \texttt{Q} is a SPARQL query that may use constructs of the form \texttt{\ourkw{QVALUES}(var')} as subqueries,  where \texttt{var'} names a solution variable. 


\paralist{Loop} Executes a sequence of statements until a termination condition holds. The syntax of a loop statement is:

\begin{center}
\texttt{\ourkw{DO} (S) \ourkw{WHILE} (condition);}
\end{center}

\noindent where \texttt{S} is a sequence of statements and \texttt{condition} is one of the following three forms of termination condition:
\begin{itemize}
\item \texttt{\ourkw{TIMES} t}, where \texttt{t} is an integer greater than 0.
\item \texttt{\ourkw{FIXPOINT} (var)}, where \texttt{var} is a solution variable. 
\item \texttt{AQ}, where \texttt{AQ} is an ASK query that may use constructs of the form \texttt{\ourkw{QVALUES}(var)} as subqueries.
\end{itemize}

\paralist{Return} Specifies the solution sequence to be returned by the procedure. The syntax of a return statement is:
\begin{center}
\texttt{\ourkw{RETURN} (var);}
\end{center}

\noindent where \texttt{var} is a solution variable. 
\medskip

Finally, a SPARQAL \textit{procedure} is a sequence of statements satisfying the following two conditions: 

\begin{itemize}
\item the last statement is a return statement and no other (nested) statement is a return statement; 
\item all solution variables used in \ourkw{QVALUES}, \ourkw{FIXPOINT} and \ourkw{RETURN} have been assigned by \ourkw{LET} in a previous statement (or a nested statement thereof).
\end{itemize}

\begin{example}
Example~\ref{ex:metro} illustrates a SPARQAL procedure with three statements, one of which contains two additional nested statements. The first statement is an assignment statement that goes from line 1 to 6. The second statement is a loop statement that goes from line 7 to 22; this statement has a \ourkw{FIXPOINT} ending condition, and it contains a sequence of two nested assignment statements: the first goes from line 8 to 14 while the second goes from line 15 to 21. The last statement, on line 23, is a return statement. \qed
\end{example}

\subsection{Semantics}

We now give the semantics of statements that form procedures in SPARQAL. More formally, let $\Pro = s_1; \dots; s_n$ be a sequence of statements, and let $\texttt{var\_1},\dots,\texttt{var\_k}$ be all variables mentioned in any statement in $\Pro$ (including in nested statements). For a tuple $\values_0 = (r_1,\dots,r_k)$ of initial assignments of (possibly empty) solution sequences to variables $\texttt{var\_1},\dots,\texttt{var\_k}$, we will construct a sequence $\values_0,\dots,\values_n$ of $k$-tuples, where each 
$\values_i$ represents the value of all variables after executing statement $s_i$. (Note that for brevity, in what follows, we assume the SPARQL dataset upon which queries are evaluated to be fixed.)

The construction is done inductively. Assume that $\values_{i-1} = (r_1,\dots,r_k)$. The value of $\values_{i}$ depends on the nature of $s_i$. 
First, if $s_i$ is the assignment statement:
\begin{center}
\texttt{\ourkw{LET} var\_j = (Q);}
\end{center}
\noindent
then tuple $\values_i$ is constructed as follows. Define SPARQL query $Q[(\texttt{var\_1},\dots,\texttt{var\_k}) \mapsto (r_1,\dots,r_k)]$ as the result of substituting each subquery 
\texttt{\{\qvalues(var\_i)\}} in $Q$ for the solution sequence $r_i$\footnote{A syntactic way of doing this is to use a \sparqlkw{VALUES} command in SPARQL.}, 
and let $r^*$ be the result of evaluating this extended query over the database. 
Tuple $\values_i$ is then defined as $$\values_i = (r_1,\dots,r_{i-1},r^*,r_{i+1},r_k),$$ \noindent that is, the result of substituting $r_i$ for $r^*$ in the 
tuple $\values_{i-1}$. 

Next, if $s_i$ is the return statement
\begin{center}
\texttt{\ourkw{RETURN}(var\_j)}
\end{center}
\noindent
Then the program terminates and returns the solution sequence $r_j$ that is the $j$-th component of $\values_i$. 

Finally, if $s_i$ is the loop statement 
\begin{center}
\texttt{\ourkw{DO} (S) \ourkw{WHILE} (condition);}
\end{center}
The tuple $\values_i$ is constructed as follows. 
Assume that $S$ is the sequence $s'_1,\dots,s'_\ell$ and notice that (by definition) $S$ must use a subset of the $k$ solution variables in $\Pro$. 
Repeat the following steps until the terminating condition is met: 

\begin{enumerate}
\item Initialize $\values'_0 \coloneq \values_{i-1}$. 
\item Compute the tuple $\values'_\ell$ that represents the result of executing statements $s'_1,\dots,s'_\ell$. 
\item If $\values'_\ell$ does not satisfy the condition, set $\values'_0 \coloneq \values'_\ell$ and repeat step 2 above. 
\item Otherwise finish, and set $\values_i \coloneq \values'_\ell$. 
\end{enumerate}

To define when a tuple $\values'_\ell$ over $k$ variables satisfies a condition, we cover all three cases: 
\begin{itemize}
\item If the condition is \ourkw{TIMES t}, then the condition is met once the loop above has repeated $t$ times. 
\item If the condition is \texttt{\ourkw{FIXPOINT} (var\_j)}, then the condition is met when the $j$-th component of $\values'_\ell$ contains the same set of solutions as the $j$-th component of $\values'_0$.
\item If the condition is \texttt{AQ}, then the condition is met when the ASK query 
$AQ[(\texttt{var\_1},\dots,\texttt{var\_k}) \mapsto \values'_\ell]$ evaluates to true. 
\end{itemize}

Note that we assume all variables to have a global scope as it makes the semantics simpler to define; one could define the semantics for variables with local scope in a similar way.

\begin{figure*}[t!]
\centering
\begin{lstlisting}
LET zika = (              # directed graph of citations between Zika articles
  SELECT ?node ?cite WHERE { 
    ?node wdt:P31 wd:Q13442814 ; wdt:P921 wd:Q202864 ; wdt:P2860 ?cite .
    ?cite wdt:P31 wd:Q13442814 ; wdt:P921 wd:Q202864 .
  }
);
LET nodes = (             # all nodes of Zika graph
  SELECT DISTINCT ?node WHERE { 
    { QVALUES(zika) } UNION { SELECT (?cite AS ?node) WHERE { QVALUES(zika) } } 
  }
);
LET n = (                 # number of nodes in Zika graph
  SELECT (COUNT(*) AS ?n) WHERE { QVALUES(nodes) } 
);
LET degree = (            # out-degree (>1) of nodes in Zika graph
  SELECT ?node (COUNT(?cite) AS ?degree) WHERE { QVALUES(zika) } GROUP BY ?node 
);
LET rank = (              # initial rank
  SELECT ?node (1.0/?n AS ?rank) WHERE { QVALUES(nodes) . QVALUES(n) } 
);
DO (                      # begin 10 iterations of PageRank
  LET rank_edge = (         # spread rank to neighbours via edges
    SELECT (?cite AS ?node) (SUM(?rank*0.85/?degree) AS ?rankEdge) WHERE { 
      QVALUES(degree) . QVALUES(rank) . QVALUES(zika)
    } GROUP BY ?cite
  );
  LET unshared = (          # compute total rank not shared via edges
    SELECT (1-SUM(?rankEdge) AS ?unshared) WHERE { QVALUES(rank_edge) }
  );
  LET rank = (              # split and add unshared rank to each node
    SELECT ?node (COALESCE(?rankEdge,0)+(?unshared/?n) AS ?rank) WHERE { 
      QVALUES(nodes) . QVALUES(n) . QVALUES(unshared) . OPTIONAL { QVALUES(rank_edge) }
    }
  );
) WHILE (TIMES 10);
LET p_index_top = (       # compute p-index for authors, select top author
  SELECT ?author (SUM(?rank) AS ?p_index) WHERE { 
    QVALUES(rank) . ?node wdt:P50 ?author .
  } GROUP BY ?author ORDER BY DESC(?p_index) LIMIT 1
);
RETURN(p_index_top);
\end{lstlisting}
\caption{Procedure to compute the top author in terms of $p$-index for articles about the Zika virus}
\label{fig:pr-zika}
\end{figure*}

\begin{example}
We recall again Example~\ref{ex:metro}, this time to illustrate the semantics of SPARQAL. In the first \ourkw{LET} statement, we assign the solution sequence of the given SPARQL query to the variable \texttt{reachable}. Then the procedure enters a loop. We assign \texttt{adjacent} to the results of a SPARQL query that embeds the current solutions of \texttt{reachable} as a sub-query, leading to a join between current \texttt{reachable} stations and pairs of adjacent stations not on Line C. We then update the \texttt{reachable} solutions, adding \texttt{adjacent} solutions; here we can use \texttt{reachable} in the \ourkw{LET} and \ourkw{QVALUES} of the same statement since it was assigned previously (line 1).  In each iteration the solutions for \texttt{reachable}  will increase, discovering new stations adjacent to previous ones, until a fixpoint. Finally, the \ourkw{RETURN} clause specifies the solutions to be given as a result of the procedure. \qed
\end{example}

\subsection{Example with PageRank}

We now illustrate a procedure for a more complex queralytic.

\begin{example}\label{ex:pr-zika}
Suppose that we have the citation network of a group of articles on a topic of interest. After obtaining such network, we want to compute a centrality algorithm in order to know which articles of the network are the most important. Thereafter we wish to use these scores to find the most prominent authors in the area. We can express this task using SPARQAL. In this case we will consider the citation network of all the articles about the Zika virus, where we then run the PageRank algorithm to know which articles are more relevant in the network, using the resulting scores to compute $p$-indexes for the respective authors. We show a procedure in our language for solving this task in Figure~\ref{fig:pr-zika}.

In this procedure we start by defining a variable that contains a solution sequence with pairs $(\texttt{?node},\ \texttt{?cite})$ such that both \texttt{?node} and \texttt{?cite} are instances of (\texttt{P31}) scientific articles (\texttt{Q13442814}) about (\texttt{P921}) the Zika virus (\texttt{Q202864}) and \texttt{?node} cites (\texttt{P2860}) \texttt{?cite}. The solutions for this query are assigned to \texttt{zika}. We can consider this variable as the representation of a directed subgraph extracted from Wikidata. We also define the variables \texttt{nodes} with all nodes in the subgraph, \texttt{n} with the number of nodes, and \texttt{degree} with the out-degree of all nodes in the graph (with some out-edge).

After extracting the graph and preparing some data structures for it, we then start the process of computing PageRank. First we assign the variable \texttt{rank} with initial ranks for all nodes of $\frac{1}{n}$. We then start a loop where we will execute 10 iterations of PageRank.\footnote{We select this termination condition for simplicity; we could also implement, for example, conditions based on residual norm, correlation coefficients, etc.} In each iteration we will first compute and assign to \texttt{rank\_edge} the PageRank that each node shares with its neighbours; here we assume a damping factor $d = 0.85$ as typical for PageRank~\cite{pagerank}, denoting the ratio of rank that a node shares with its neighbours. Next we compute and assign to \texttt{unshared} the total rank not shared with neighbours in the previous step (this arises from nodes with no out-edges and the $1-d$ factor not used previously for other nodes). We then conclude the iteration by splitting and adding the unshared rank to each node equally, updating the results for \texttt{rank}. The loop is applied 10 times, computing PageRank for each article.

Finally, we join the PageRank scores for articles with their authors, and use an aggregation to sum the scores for each author, applying ordering and a limit to select the top author, assigning the solution to \texttt{p\_index\_top}. 
Finally, the procedure returns the solution for \texttt{p\_index\_top} denoting the top author. \qed
\end{example}

\section{Evaluation in Batches}\label{sec:batch}

%
%
%

Procedures in SPARQAL use {\ourkw{QVALUES}} clauses to coordinate solution sequences between statements, allowing to pass, extend and refine data throughout the procedure. A natural way to coordination solution sequences across statements is to store them in memory during the execution of the procedure; however, large solution sequences may not fit in memory. 
To alleviate this issue, we develop an alternative approach to perform the joins instigated by \texttt{\ourkw{QVALUES}} clauses in batches, using a technique reminiscent of the Map-Reduce paradigm.\footnote{The approach is also similar to ``shipping strategies'' for federated queries~\cite{ArandaPU14}.} This approach allows to evaluate queries without assuming that intermediate solution sequences fit in memory and, moreover, allows to parallelise the evaluation of queries.

\subsection{Overall Strategy}
The strategy for evaluation in batches is as follows. First, each SPARQL query in a (nested) statement of the procedure is associated with \texttt{Map} and \texttt{Reduce} functions. These functions replace a query $Q$ working with one or more \texttt{\ourkw{QVALUES}(var)} clauses -- typically evaluated in full and passed to the query -- to a sequence of queries $Q_1,\dots,Q_n$ in which the instantiations of \texttt{\ourkw{QVALUES}(var)} clauses only retrieves a subset of the tuples in variable \texttt{var}. These queries -- representing batches -- are generated by the \texttt{Map} function. The \texttt{Reduce} function then merges the results for each $Q_1,\dots,Q_n$ into a single output. Because these queries are evaluated separately, and over smaller portions of the solution sequence, this approach reduces memory requirements and enables parallel evaluation. The downside is that we now execute a series of queries, instead of one. 

Before formally defining the strategy, we provide an example.

\begin{example}\label{ex:batch}
Recall Example \ref{ex:pr-zika} and the procedure to compute the top author in terms of $p$-index for Zika articles. Consider the example solution sequence for the variable \texttt{zika} shown in Figure~\ref{fig:zika-data} alongside the directed graph it represents (in practice, Wikidata returns over 3 thousand articles with over 38 thousand citations).

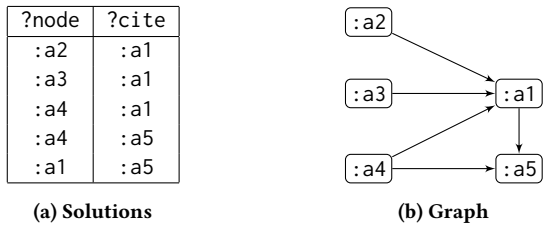
\begin{figure}
	\begin{subfigure}[b]{0.45\columnwidth}
	    \centering
		\begin{tabular}{ |c|c| } 
 			\hline
			\texttt{?node} & \texttt{?cite} \\ 
			\hline
			\texttt{:a2} & \texttt{:a1} \\ 
			\texttt{:a3} & \texttt{:a1} \\
			\texttt{:a4} & \texttt{:a1} \\ 
			\texttt{:a4} & \texttt{:a5} \\
			\texttt{:a1} & \texttt{:a5} \\
			\hline
		\end{tabular}
		\caption{Solutions}	
		\label{tab:cit-net}
	\end{subfigure}
	\hfill
	\begin{subfigure}[b]{0.45\columnwidth}
	    \centering
		\begin{tikzpicture}
		\tikzset{vertex/.style = {shape=rectangle,rounded corners=2pt,draw,minimum size=1.2em,inner sep=2pt}}
		\tikzset{edge/.style = {->,> = latex'}}
		
		\node[vertex] (a) at  (0,0) {\texttt{:a1}};
		\node[vertex] (b) at  (-2,0.95) {\texttt{:a2}};
		\node[vertex] (c) at  (-2,0) {\texttt{:a3}};
		\node[vertex] (d) at  (-2,-1) {\texttt{:a4}};
		\node[vertex] (e) at (0,-1) {\texttt{:a5}};
		
		\draw[edge] (b) to (a);
		\draw[edge] (c) to (a);
		\draw[edge] (d) to (a);
		\draw[edge] (d) to (e);
		\draw[edge] (a) to (e);
		\end{tikzpicture}
		\caption{Graph}
		\label{fig:cit-net}
	\end{subfigure}
	\caption{Example results for \texttt{zika} solution variable}\label{fig:zika-data}
\end{figure}

Next consider the assignment of the variable \texttt{rank\_edge} on line~22 at the first iteration of the loop. Intuitively, this assignment computes how much PageRank score each article will receive from its citations. Instead of evaluating the query as usual, we will use a \texttt{Map} function in order to evaluate it in several batches. More specifically, let $Q_{\texttt{rank\_edge}}$ be the query that assigns the variable \texttt{rank\_edge}. We associate $Q_{\texttt{rank\_edge}}$ with the following 
\texttt{Map} and \texttt{Reduce} functions. 
Our \texttt{Map} function receives two inputs: a SPARQL variable \texttt{?v} and a unary \texttt{SELECT} SPARQL query that mentions \texttt{?v}. This corresponds to 
the invocation \texttt{Map(\texttt{?cite}, [$Q_{\texttt{node}}$])} in our notation, where $Q_{\texttt{node}}$ is the following query that assigns to \texttt{?node} all articles that cite 
the article assigned to the query variable \texttt{?cite}. 

\medskip
\begin{lstlisting}
SELECT ?node WHERE { 
  ?node wdt:P31 wd:Q13442814 ; 
    wdt:P921 wd:Q202864 ; wdt:P2860 ?cite .
}
\end{lstlisting}

%
%


This \texttt{Map} function will divide the query $Q_{\texttt{rank\_edge}}$ into a series of queries: one for each node of the citation network; this is done by splitting the solution sequence of the variable \texttt{zika} into a set of sequences where the binding of variable \texttt{?cite} is different. In our case, this corresponds to elements \texttt{:a1} and \texttt{:a5}. Thus, by splitting variable \texttt{zika} into two variables \texttt{zika\_a1} and \texttt{zika\_a5} -- each of them instantiated with 
the respective solution sequence -- we define two 
different queries for $Q_{\texttt{rank\_edge}}$: the first invokes \texttt{\ourkw{QVALUES}(\texttt{zika\_a1})} and the second invokes \texttt{\ourkw{QVALUES}(\texttt{zika\_a5})}.


We call these queries $Q_{\texttt{:a1}}$ and $Q_{\texttt{:a5}}$. Intuitively, they are meant to compute the result of $Q_{\texttt{rank\_edge}}$ in two different batches. 
Let us start with query $Q_{\texttt{:a1}}$. As mentioned, this query excludes all the tuples of the solution sequence stored in the variable \texttt{zika} where the value of \texttt{?cite} is not \texttt{:a1}. This implies that the \texttt{\ourkw{QVALUES}(zika)} clause should be replaced by the solution sequence (batch) shown in Figure~\ref{tab:sq0} labelled \texttt{zika\_a1}; here, the mappings with the values \texttt{(:a4,:a5)} and \texttt{(:a1,:a5)} are not considered (they form \texttt{zika\_a5}). 

Now we need to assign solution sequences to the variables \texttt{rank} and \texttt{degree} corresponding to $Q_{\texttt{:a1}}$ (and later $Q_{\texttt{:a5}}$, respectively). While we could assign the full solution sequences to these variables, this would defeat the purposes of batching and is unnecessary: to compute the PageRank of (e.g.) the node \texttt{:a1} we only need information about the neighbours of \texttt{:a1}, not the entire graph. Instead, we again split \texttt{rank} and \texttt{degree}, making use of the query $Q_{\texttt{node}}$ in the definition of \texttt{Map}: we define one extra variable \texttt{?node}, and we evaluate a copy of query $Q_{\texttt{node}}$ in which the variable \texttt{?cite} is replaced by \texttt{:a1}, thus effectively storing in \texttt{?node} all papers that cite \texttt{:a1}. We use these values, and filter out 
any solution sequence of \texttt{degree} that is not binding \texttt{?node} to one of these values. In this case the \texttt{\ourkw{QVALUES}(degree)} is replaced by the solution sequence shown in Figure~\ref{tab:sq1}. We do the same for the \texttt{\ourkw{QVALUES}(rank)} clause, which is replaced by the solution sequence shown in Figure~\ref{tab:sq2}.

\begin{figure}
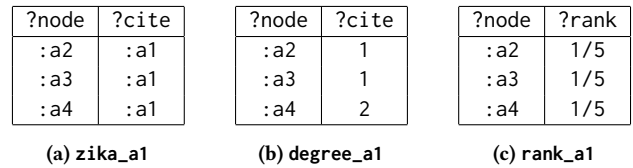

	\begin{subfigure}[b]{0.3\columnwidth}
	    \centering
		\begin{tabular}{ |c|c| } 
		 \hline
		 \texttt{?node} & \texttt{?cite} \\ 
		 \hline
		 \texttt{:a2} & \texttt{:a1} \\ 
		 \texttt{:a3} & \texttt{:a1} \\
		 \texttt{:a4} & \texttt{:a1} \\ 
		 \hline
		\end{tabular}
		\caption{\texttt{zika\_a1}}
		\label{tab:sq0}
	\end{subfigure}
	\hfill
	\begin{subfigure}[b]{0.3\columnwidth}
	    \centering
		\begin{tabular}{ |c|c| } 
		 \hline
		 \texttt{?node} & \texttt{?cite} \\ 
		 \hline
		 \texttt{:a2} & \texttt{1} \\ 
		 \texttt{:a3} & \texttt{1} \\
		 \texttt{:a4} & \texttt{2} \\ 
		 \hline
		\end{tabular}
		\caption{\texttt{degree\_a1}}
		\label{tab:sq1}
	\end{subfigure}
	\hfill
	\begin{subfigure}[b]{0.3\columnwidth}
	    \centering
		\begin{tabular}{ |c|c| } 
		 \hline
		 \texttt{?node} & \texttt{?rank} \\ 
		 \hline
		 \texttt{:a2} & \texttt{1/5} \\ 
		 \texttt{:a3} & \texttt{1/5} \\
		 \texttt{:a4} & \texttt{1/5} \\ 
		 \hline
		\end{tabular}
		\caption{\texttt{rank\_a1}}
		\label{tab:sq2}
	\end{subfigure}
	\caption{Intermediate solution sequences for \texttt{:a1} batch}\label{fig:zika-sols}
\end{figure}

Now if we evaluate the query $Q_{\texttt{:a1}}$ replacing the \ourkw{QVALUES} clauses with the respective batches of solution sequences, we would, in turn, obtain the solution sequence \texttt{\{(:a1 0.425)\}}. If we repeat the process for \texttt{:a5}, the query $Q_{\texttt{:a5}}$ results in the solution sequence \texttt{\{(:a5 0.255)\}}. Since we need to create a single solution sequence to assign to the variable \texttt{rank\_next}, we now use a \texttt{Reduce} function; in this case, we will simply take the \sparqlkw{UNION} of the batched solution sequences. The result is then the same as we would have obtained by evaluating the full solution sequences each time. \qed
\end{example}

This strategy of batching solution sequences thus reduces memory requirements. Note that a process like this could be continued for every query extended with \ourkw{QVALUES} in the procedure of Example~\ref{ex:pr-zika}. We now formally define the strategy.

\subsection{Formal Definition}

The strategy we presented has two steps. The first one is the \texttt{Map} step, where we define how to split solution sequences, and the second is the \texttt{Reduce} step, where 
we group together the batches we evaluate. To formally define how these operators work, we will assume that we are writing \texttt{Map} and \texttt{Reduce} steps for a 
query $Q$ that uses clauses \texttt{\ourkw{QVALUES}(var\_1)} $,\dots,$ \texttt{\ourkw{QVALUES}(var\_k)}.  

\paralist{Map} The \texttt{Map} operator has the following syntax:
$$
\texttt{Map}(?v, [I_1, \dots, I_m]))
$$

\noindent
where $?v$ is a SPARQL variable and $I_1, \dots, I_m$ are unary standard SPARQL queries, that is, queries that project only one variable. We assume 
each query $I_j$ projects the variable $?v_{I_j}$ for $j \in [1,\dots,m]$.

Let us assume that 
upon calling query $Q$, each clause of the form \texttt{\ourkw{QVALUES}(var\_i)} is instantiated with a solution sequence $r_i$, for $i \in [1,\dots,k]$, and define the 
set  $QDom(Q, ?v)$ as the union of all values bound to the SPARQL variable $?v$ in any of the sequences $r_1,\dots,r_n$; that is, 
if we use $r[?v]$ to denote the set of all elements that are bound to $?v$ in any mapping in $r$, then 
$$QDom(Q, ?v) = r_1[?v] \cup \dots \cup r_n[?v].$$ 

The output of the Map function is a set of tuples of solution variables of the form (\texttt{var\_c\_1},\dots,\texttt{var\_c\_k}), for $c \in QDom(Q, ?v)$, each of which stores a solution sequence $r_{c,i}$ ($i \in [1,\dots,k]$).

Let us use $I_j[?v \rightarrow c]$ to denote the SPARQL query $I_j$ where all appearances of variable $?v$ are replaced with value $c$. 
For every value $c \in QDom(Q, ?v)$ and solution sequence $r_i$, $i \in [1,\dots,k]$, we define $r_{c,i}$ as the subset of $r_i$ satisfying the following conditions.  
\begin{itemize}
\item If there is at least one mapping in $r_i$ that binds variable $?v$, then $r_{i,c}$ contains exactly those mappings in $r_i$ that bind $?v$ to value $c$. 
\item Otherwise $r_{i,c}$ contains all mappings that bind any of the variables $?v_{I_j}$ to the result of the query $I_j[?v \rightarrow c]$, respectively, for $j \in [1,\dots,m]$
\item If $r_i$ does not contain a mapping that binds $?v$ or any of $?v_{I_1},\dots,?v_{I_m}$, then $r_{i,c} = r_i$. 
\end{itemize}

Note that we are defining the \texttt{Map} function in terms of a single variable $?v$, but it is possible to extend our definition to a set of variables $?v_1,\dots,?v_n$. In this case we should consider tuples $(c_1,\dots,c_n) \in QDom(Q, ?v_1) \times \dots \times QDom(Q, ?v_n)$.

\paralist{Reduce} The \texttt{Reduce} function specifies how solution sequences are merged together; it can be their union, the sum of all bindings for a variable in their union, their intersection, etc. Each reducer receives one of the tuples (\texttt{var\_c\_1},\dots,\texttt{var\_c\_k}), for $c \in QDom(Q, ?v)$, 
each of which stores a solution sequence $r_{c,i}$, $i \in [1,\dots,k]$. With these variables, it evaluates the query $Q_c$, which results in 
replacing every instance of a construct \texttt{\ourkw{QVALUES}(var\_i)}, with \texttt{\ourkw{QVALUES}(var\_c\_i)}, for $i \in [1,\dots,k]$. 
Once all queries have been evaluated by each reducer, all intermediate results of queries $\{Q_c \mid c \in QDom(Q, ?v)\}$ are merged together 
per the \texttt{Reduce} function (in Example \ref{ex:batch}, the \texttt{Reduce} function just computes the union of all sequences).

\section{Expressive Power}

In this section we review the expressive power of procedures in SPARQAL. Our results come in two flavours: first we focus on what the language can do, showing Turing-completeness and complexity results, and then we turn to the comparison between our language and other related query languages extended with recursion. 

\subsection{Turing-completeness}

Although \texttt{do-while} loops may appear to be just a mild extension to a query language, our first result states that this is actually enough to achieve Turing-completeness. Formally, we say that a query language $\mathcal L$ is Turing-complete if for every Turing machine $M$ over an alphabet $\Sigma$ one can construct a query $Q$ in $\mathcal L$ and define a computable function $f$ that takes a word in $\Sigma^*$ and produces an RDF graph, and such that a word $w \in \Sigma^*$ is accepted by $M$ if and only if the evaluation of $Q$ over graph $f(w)$ produces a non-empty result. Along these lines, we prove the following result: 

\begin{theorem}
\label{teo:tc}
SPARQAL is Turing-complete
\end{theorem}

The proof of this theorem (presented in the extended version of this paper~\cite{online-appendix}) relies on the combination of \texttt{do-while} loops and the ability to create new values in the base SPARQL language through \sparqlkw{BIND} statements and algebraic functions~\cite{sparql11}. Of course, for the proof one must assume that there is no limit on the memory used by the evaluation algorithm; however, the proof reveals a linear correspondence between the memory used by the query and the number of cells visited by the machine $M$. 

Traditional theoretical results have tended to study languages assuming that the creation of new values is not possible, or, if possible, that there is a bound on the number of values that are created. But this is not the case with SPARQAL procedures; for starters, we can iterate and sum to create arbitrarily big numbers. However, for the purpose of comparing SPARQAL procedures against other traditional database languages, we ask, what would be its expressive power if one disallows the creation of new values? In fact, \texttt{do-while} loops have been studied previously in the literature, especially in the context of relational algebra (see e.g. \cite{AHV95}). In our context, we ask what happens if we disallow the invention of new values in the procedure: more formally, we say that a procedure $P$ \emph{does not invent new values} if for every graph $G$ and every variable \texttt{var} defined in $P$, all mappings in 
any solution sequence associated to \texttt{var} always binds variables to values already present in $G$. In this case, there is a limit on the maximum number of mappings in the solution sequence of any variable at any point in time during evaluation of the procedure, and this limit depends polynomially on the size of the graph. This implies that the evaluation of this procedure can be performed in PSPACE (in data complexity), and we can also show that this bound is tight. To formally state this result, let $P$ be a SPARQAL procedure. The evaluation problem for $P$ receives a graph as an input, and asks whether the evaluation of $P$ over $G$ is not empty.\footnote{This corresponds to boolean evaluation. This is without loss of generality because the standard evaluation problem where one considers a tuple of values as an input can be simulated by means of filters.} We can then state the following:

\begin{proposition}
\label{prop-complexity}
The evaluation problem for SPARQAL procedures that do not invent new values is \text{PSPACE}-complete. 
\end{proposition}

\subsection{Comparison with Similar Languages}

We now turn to the comparison between our language and similar proposals in the literature. 

\paralist{Recursive extensions to SPARQL} We base our comparison on the recursive extension proposed by Reutter et al.~\cite{ReutterSV15}, but these results apply to similar languages, such as the (with) recursive operator in SQL. The first observation is that these languages only define semantics for monotone queries. For example, recursive SPARQL uses constructs of the form: 

\begin{lstlisting}
WITH RECURSIVE G AS {$Q_\text{CONSTRUCT}$}
$Q_\text{SELECT}$
\end{lstlisting}

\noindent
where $G$ is an IRI used to denote a temporal graph, $Q_\text{CONSTRUCT}$ is a CONSTRUCT SPARQL query and $Q_\text{SELECT}$ is a SELECT SPARQL query. The idea of this form of recursion is that $Q_\text{CONSTRUCT}$ defines a query meant to compute $G$ in an iterative fashion (there may also be reference to the graph $G$ inside this same query). 
In other words, we can view $Q_\text{CONSTRUCT}$ as an operator 
$T_Q(G)$ that -- as a single step -- takes as input an RDF graph and produces as output an RDF graph. The final output graph then corresponds to the least fixed point of the sequence 
$T_Q(\emptyset)$, $T_Q(T_Q(\emptyset))$, $\dots$. Such a fixed point is only guaranteed when $Q_\text{CONSTRUCT}$ is \textit{monotone}: where  
$G \subseteq G'$ implies that $T_Q(G) \subseteq T_Q(G')$. To guarantee having monotone queries, Reutter et al.~\cite{ReutterSV15} impose major syntactic restrictions on the operands available for the $Q_\text{CONSTRUCT}$ query, forbidding, for example, the use of \sparqlkw{BIND}, \sparqlkw{NOT EXISTS}, \sparqlkw{MINUS}, as well as \sparqlkw{OPTIONAL} patterns that are not \textit{well designed}~\cite{perez2009semantics}.

So how does our language compare with these recursive variants? The first observation is that all of these queries can actually be expressed as a SPARQAL procedure: 
a query in the form above can be straightforwardly simulated by the following procedure:  
\begin{lstlisting}
DO (
  LET graph = (
    SELECT ?s ?p ?o WHERE $P'_\text{CONSTRUCT}$	
  )
) WHILE ( FIXPOINT (graph) )
LET result = $Q'_\text{SELECT}$;
RETURN result;
\end{lstlisting}
Here $P'_\text{CONSTRUCT}$ is the pattern corresponding to the \sparqlkw{WHERE} part of $Q_\text{CONSTRUCT}$ from the recursive SPARQL query, but where instead of using temporal graph $G$ we retrieve those triples from 
the subquery \texttt{\ourkw{QVALUES}(graph)}. Query $Q'_\text{SELECT}$ corresponds to $Q_\text{SELECT}$ from the recursive SPARQL query, but where again we use \texttt{\ourkw{QVALUES}(graph)} instead of the temporal graph $G$. 

In the other direction, can recursive SPARQL simulate SPARQAL procedures? This depends on 
what sort of queries we allow in $Q_\text{CONSTRUCT}$. If we take the language as originally defined by Reutter et al., 
so that queries $Q_\text{CONSTRUCT}$ must be monotone, then we know that the evaluation for 
recursive SPARQL queries is in PTIME \cite{ReutterSV15}. Together with Proposition \ref{prop-complexity}, 
this means that recursive SPARQL cannot simulate SPARQAL procedures unless PTIME = PSPACE, which is widely assumed to be false. 
We also remark that a similar result was shown for similar extensions to relational algebra: relational algebra equipped with fixed point cannot 
simulate do-while queries unless PTIME = PSPACE \cite{AHV95}.

On the other hand, when one allows to use operands such as \sparqlkw{BIND} clauses, 
the operator given by $Q_\text{CONSTRUCT}$ becomes non-monotone, and the semantics for this case is not defined. 
The standard solution for this case is to assign a partial fixed point semantics, which means that a query of the form above would retrieve a graph $G$ which is the fixed point of the sequence 
$T_Q(\emptyset)$, $T_Q(T_Q(\emptyset))$, $\dots$, if it exists, or an empty graph otherwise (when the operator runs into an infinite loop). In this context, and if we allow full SPARQL 1.1 in $Q_\text{CONSTRUCT}$, 
one can actually show that both languages coincide, because recursive SPARQL becomes Turing-complete as well. 

\paralist{Graph Neural Networks (GNNs)} Another framework for graph analytics 
that has recently received considerable attention is that of GNNs (see e.g. \cite{battaglia}). Roughly speaking, the basic architecture for GNNs 
consists of a sequence of $L$ \emph{layers} that combine the feature vectors $\vx_v$ of every node $v$ of the graph with the multiset of feature vectors of its neighbours. 
Formally, let $\mathcal{N}_G(v)$ contain all neighbours of a node $v$ in $G$. 
For each layer one defines sets of aggregation and combination functions $\{\AGG^{(i)}\}_{i=1}^L$ and $\{\COM^{(i)}\}_{i=1}^L$, 
and vectors $\vx_v^{(i)}$ of graph labels are computed for every node $v$ of a graph $G$ via the following recursive formula, 
for $i = 1,\dots,L$:
\begin{equation}\label{eq:ac-gnn}
\vx_v^{(i)} = \COM^{(i)}\bigg(\vx_v^{(i-1)}, \AGG^{(i)}\big(\lms \vx_u^{(i-1)}\mid u\in \mathcal{N}_G(v)\rms\big)\bigg) 
\end{equation}

\noindent
where each $\vx_v^{(0)}$ is the initial feature vector $\vx_v$ of $v$. 
GNNs also assume a final classification or readout functions to compute a global vector for the graph, that is applied at the end of 
the computation. 

Thus, in terms of graph analytics, GNNs can be seen as functions that receive a graph as an input, and output either a global value or 
another graph that has the same nodes and edges, but where the 
label of nodes (and, in full generality, edges) may have been modified. We remark that this framework is congruous with the systolic abstraction at the heart of various frameworks for graph analytics~\cite{DeLorimierKMRERUKD06,MalewiczABDHLC10,KrepskaKFB11,GonzalezLGBG12,XinGFS13,ChingEKLM15,signalcollect}, as discussed previously.

It is thus of interest to compare GNNs to our SPARQAL language; for this, we assume that we deal with RDF graphs in which all nodes 
are assigned a label via a triple with the property \texttt{rdfs:label}. Of course, since SPARQAL procedures are Turing-complete, one can simulate any GNN with such a procedure. What is more interesting to study is to reverse the question: 
to understand how GNNs relate to the expressive power of restricted forms of SPARQAL. 

As previously mentioned, it was recently shown~\cite{XuHLJ19,morris} that the power of GNNs in terms of computing vectors of nodes is bounded by, and captures, the 
Weisfeiler--Lehman (WL) graph isomorphism test~\cite{CFI92}. 
The WL test can be understood as a procedure that starts from a labelled graph, and iteratively assigns, for a certain number of \emph{rounds}, 
a new label to every node in the graph; this is done in such a way that the label of a node in each round has a one-to-one correspondence with 
its own label and the multiset of labels of its neighbours in the previous round. If the WL test on a given graph $G$ assigns the same label 
to two nodes $a$ and $b$ of $G$, then every GNN must also assign the same label to both of these nodes~\cite{XuHLJ19,morris}: this is because GNNs can only aggregate local information for nodes.


In what follows we will define a restricted form of procedure in SPARQAL whose expressive power is comparable to that of GNNs, i.e., that it is bounded by, and captures, the WL test. Formally, we define a \emph{local SPARQAL procedure} as a procedure of the form:

\begin{lstlisting}
LET var_1 = ($Q_1$); 
$\vdots$
LET var_k = ($Q_k$);
LET vector = (
  SELECT ?v ?lab WHERE { ?v rdfs:label ?lab }); 
DO ( 
  $\vdots$
) WHILE ( condition );
RETURN(vector);
\end{lstlisting}

\noindent
such that (i) each query $Q_1,\dots,Q_k$ is a basic graph pattern of the form $\{?v\ p_j\ ?v_j\}$ or $\{?v_j\ p_j\ ?v\}$, for variables $?v, ?v_1,\dots,?v_k$ and properties 
$p_1,\dots,p_k$; (ii) all statements in the \texttt{do-while} loop only use variables $?lab,?v, ?v_1,\dots,?v_k$ in their queries, and no constants 
(that is, they cannot retrieve any further information from the graph), 
and (iii) queries in the \texttt{DO-WHILE} loop are evaluated in the \texttt{Map}/\texttt{Reduce} framework, but where the \texttt{Map} function is just 
$\texttt{Map}(?v)$. The intuition behind this is as follows. Solution variables \texttt{var\_1},\dots,\texttt{var\_k} are restricted so that all they can store are tuple of values describing parts of the neighbourhood of a 
node. Then, the iteration can only look at this neighbourhood, and update the label according to this information. We now state our result. 

\begin{theorem}
\label{theo-gnns}
The power of local SPARQAL procedures is bounded by, and captures, the WL-test; specifically: 
\begin{itemize}
\item When running any local SPARQAL procedure over a graph $G$, if there are nodes $a$ and $b$ that are assigned the same label by the WL-test, then 
the returned sequence $r$ for variable \texttt{vector} must be such that for any two mappings $\mu_1$ and $\mu_2$ in $s$ where $\mu_1[?v] = a$ and $\mu_2[?v] = b$, 
it holds that $\mu_1[?lab] = \mu_2[?lab]$. 
\item There is a local SPARQAL procedure $P$ that can reproduce the WL test: for every graph $G$, the output of $P$ over $G$ is the same as the 
output of the WL test over $G$. 
\end{itemize}
\end{theorem}

Together with the result that GNN are also bounded, and capture, the WL-test \cite{XuHLJ19,morris}, we have that local SPARQAL procedures are comparable in term of expressivity to GNNs.

\section{Experiments}

In this section we present our implementation of a \querylytics\ engine based on the SPARQAL language. This implementation was developed on top of the Apache Jena Framework, version 3.10. The core implementation provides the following core functionalities: (1) it parses the SPARQAL procedure into a sequence of statements, which are evaluated according to their semantics by: (2a) maintaining a map where the key is the variable name and the value is the solution sequence; (2b) replacing variables used within a \texttt{\ourkw{QVALUES}} clause with a \texttt{\sparqlkw{VALUES}} string with the respective solution sequence; (2c) evaluating SPARQL queries, and (2d) in order to handle \texttt{\ourkw{FIXPOINT}} conditions, maintaining the previous solution sequence of the respective variable in-memory to monitor changes. We further implement the Map/Reduce strategy defined in Section~\ref{sec:batch}.

We adopt a query engine for the current implementation as our target use-case  is -- per the scenarios outlined in Examples~\ref{ex:metro} and~\ref{ex:pr-zika} -- to run queralytics (near-)interactively on small-to-medium size graphs that have been projected from a larger graph using a query. We first report results for our two motivating scenarios. We then devise a benchmark based on Wikidata for running popular analytical tasks on selective sub-graphs that are similarly extracted through queries. Finally, though not part of our target use-case, we stress-test our implementation for a graph analytics benchmark at a larger scale, including results for the Map/Reduce framework designed to reduce memory requirements by using batches.

Experiments were tested on a MacBook Pro with a 3.1 GHz Intel I5 processor and 16 GB of RAM. The source code, procedures and datasets used are available online in an anonymous appendix \cite{online-appendix}.


\subsection{Wikidata: Motivating Examples}

Our first experiment is to anecdotally evaluate the procedures described in Examples~\ref{ex:metro} and~\ref{ex:pr-zika}, evaluating the Buenos Aires metro and Zika $p$-index \querylytics. Example \ref{ex:metro} took just 1.3 seconds to return 16 stations from which Palermo can be reached without using Line C. Example \ref{ex:pr-zika} -- running 10 iterations of PageRank on a graph of 38,738 edges (citations) and 3,057 nodes (articles)  -- took 53.1 seconds to find the top author (from 2,214 authors) according to their $p$-index in the citation network; for reference, Table~\ref{tab:zika-res} shows the results for the top 5 authors, ordered by their $p$-index.

\begin{table}[t]
\centering
\caption{Top-5 authors according to their Zika $p$-index\label{tab:zika-res}}
\vspace{-1em}
\begin{tabular}{ |c|c|c| } 
 \hline
 \texttt{?author} & \texttt{?p\_index} & \texttt{?name} \\ 
 \hline
 \texttt{wd:Q18876341} & \texttt{0.124} & \texttt{George Dick} \\ 
 \texttt{wd:Q24696365} & \texttt{0.084} & \texttt{Ademola H. Fagbami} \\ 
 \texttt{wd:Q21165078} & \texttt{0.083} & \texttt{Alexander John Haddow} \\ 
 \texttt{wd:Q24515005} & \texttt{0.078} & \texttt{Stuart Fordyce Kitchen} \\ 
 \texttt{wd:Q24727761} & \texttt{0.046} & \texttt{Robert S. Lanciotti} \\ 
 \hline
\end{tabular}

\end{table}

\subsection{Wikidata: Queralytics Benchmark}

To the best of our knowledge, there is no existing benchmark for queralytics along the lines discussed in this paper (and exemplified by the previous motivating examples). This led us to devise a novel benchmark for queralytics on the Wikidata knowledge graph. We took the ``truthy'' RDF dump of Wikidata as our benchmark graph~\cite{MalyshevKGGB18}. Designing the queralytic tasks required collecting and combining two elements: queries that return results corresponding to graphs, and graph algorithms to apply analytics on these graphs. 

In terms of the queries returning graphs, we revised the list of use-case queries for the Wikidata Query Service\footnote{\url{https://www.wikidata.org/wiki/Wikidata:SPARQL_query_service/queries/examples}}. From this list, we identified the following six queries returning graphs:

\begin{description}
\item[Q1] A graph of adjacent metro stations in Buenos Aires
\item[Q2] A graph of citations for articles about the Zika virus
\item[Q3] A graph of characters in the Marvel universe and the groups they belong to
\item[Q4] A graph of firearm cartridges and the cartridges they are based on
\item[Q5] A graph of horses and their lineage
\item[Q6] A graph of drug--disease interactions on infectious diseases
\end{description}

These queries provide a mix of connected graphs, disconnected graphs, bipartite graphs, trees, DAGs, near-DAGs, and so forth. We provide the sizes of these graphs in Table~\ref{tab:qgraphs}, where we see that the smallest graph is indeed the Buenos Aires metro graph, while the largest is the citation graph for Zika articles.

\begin{table}[t]
\caption{Number of nodes and edges in graphs considered \label{tab:qgraphs}} 
\vspace{-1em}
\begin{tabular}{|l|r|r|}
\hline
\textbf{Id} & \textbf{Nodes} & \textbf{Edges} \\
 \hline\hline
\textbf{Q1} & 93 & 172 \\
\textbf{Q2} & 3,057 & 38,738 \\
\textbf{Q3} & 480 & 766 \\
\textbf{Q4} & 266 & 211 \\
\textbf{Q5} & 7,194 & 8,719 \\
\textbf{Q6} & 627 & 996 \\
\hline
\end{tabular}
\end{table}

Next we must define the analytics that we would like to apply on these graphs. For this, we adopted five of the six algorithm sproposed for the Graphalytics Benchmark~\cite{graphalytics} defined by the Linked Data Benchmark Council (LDBC); namely:

\begin{description}
\item[BFS] Breadth-First Search
\item[LCC] Local Clustering Coefficient
\item[PR] PageRank
\item[SSSP] Single-Source Shortest Path
\item[WCC] Weakly Connected Components
\end{description}

\noindent
We do not include the \textit{Community Detection through Label Propagation} algorithm as not all our graphs 
have natural categorical labels upon which this analytical task depends (we will test this algorithm in the experiment that follows, however). We implement these five algorithms as procedures in the SPARQAL language, prefixing each with the six different Wikidata graph queries, stored as solution variables. The result is a benchmark of $6 \times 5 = 30$ queralytic tasks. 

\begin{figure}
\includegraphics[scale=0.85]{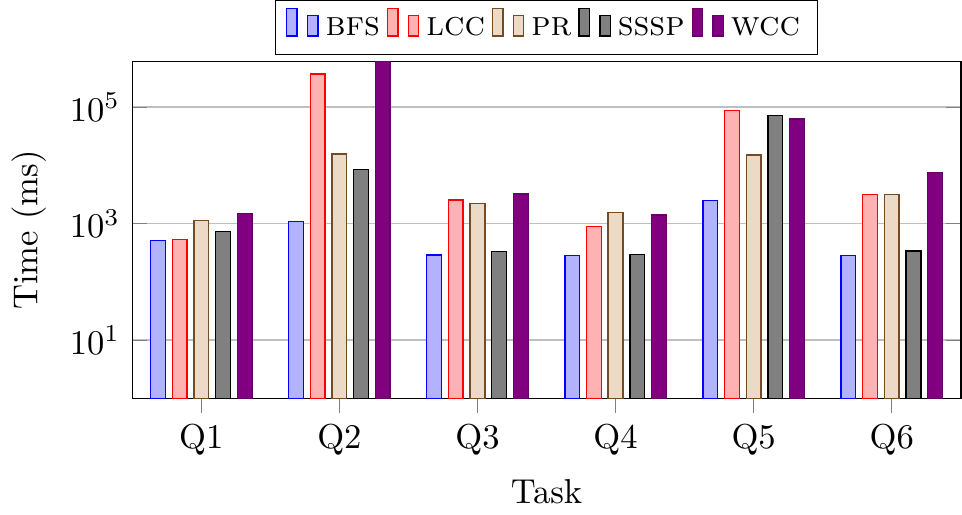}
\caption{Result for Wikidata queralytic benchmark}\label{fig:wikiresults}
\end{figure}

In Figure~\ref{fig:wikiresults}, we show the results for these 30 tasks using our in-memory implementation. First we remark that the Weakly Connected Components (WCC) algorithm timed-out in the case of the Zika graph after 10 minutes; furthermore, the LCC algorithm failed with memory errors on the Zika graph, where the time shown is thus for the Map/Reduce implementation. While the cheapest algorithm in general was BFS, the most expensive was WCC.  Although some of these tasks took over a minute in the case of graphs with thousands or tens of thousands of nodes (Zika/Q1 and Horses/Q5), those with fewer than a thousand nodes/edges ran in under a second, compatible with interactive use. 

\subsection{Graphalytics: Stress Test}

The scale of the previous graphs is quite low and uses (mostly) the in-memory algorithm. Hence we use the Graphalytics Benchmark~\cite{graphalytics} to perform stress tests for our implementation at larger scale. We adopt the \texttt{cit-Patents} dataset: a directed graph with 3,774,768 vertices and 16,518,947 edges. We implement SPARQAL procedures to run six graph algorithms on the full graph; in particular, we run the aforementioned five algorithms, as well as:

\begin{description}
\item[CDLP] Community Detection through Label Propagation
\end{description}

The results of the Graphalytics benchmark are shown in Table~\ref{tab:querylytics-suck} using the in-memory algorithm; for comparison, we also offer the times using an in-memory Python implementation. We see that the results are overwhelmingly negative, with poor performance due in particular to our handling of \texttt{\ourkw{QVALUES}} clauses, which leads to unwieldy query strings when replaced by \texttt{\ourkw{VALUES}} for large solution sequences. Switching to the Map/Reduce approach only solved half of our problems: although the procedures did not fail, they took even longer than the in-memory cases, where in other cases we estimated that the procedure would take months to finish due to the number of queries generated.

\begin{table}[t]
\caption{Execution time (min) for Graphalytics}
\label{tab:querylytics-suck}
\vspace{-1em}
\centering
\begin{tabular}{ |l|r|r| } 
 \hline
 \textbf{Algorithm} & \textbf{SPARQAL/Jena}  &  \textbf{Python} \\ 
 \hline
 \textbf{BFS} & 11 & 1 \\ 
 \textbf{CDLP} & out of mem & 15 \\
 \textbf{LCC} & out of mem & 2 \\ 
 \textbf{PR} & 250 & 5 \\
 \textbf{SSSP} & 300 & 1 \\ 
 \textbf{WCC} & out of mem & 1 \\ 
 \hline
\end{tabular}
\end{table}

These results clearly demonstrate the limitations of our Jena-based implementation for large-scale graphs. While this is not currently our focus -- which is rather achieving interactive performance on small-to-medium graphs -- we identify this as an interesting challenge: can procedures in SPARQAL be optimised enough to be competitive with the imperative Python times shown?

\section{Conclusion}

We propose a declarative language called SPARQAL that allows for interleaving queries and analytics on graphs. We see this language as being useful in applications where analytical tasks require complex pre--and post--processing of the graph and results. In this context, we have proven some formal properties for our language, and discussed its formal relation to similar languages and abstractions. We have also implemented an initial system to support our language based on an off-the-shelf SPARQL query engine, showing that it offers interactive runtimes for typical analytics on graphs of fewer than one-thousand nodes (generated by means of a query). On the other hand, there is still much work to do if one wants a system supporting a declarative language that is competitive with standard frameworks for graph analytics. In particular, we need to look at the problem of how to compile and optimise SPARQAL procedures, ideally into smaller, lower-level components that can be implemented within database engines or analytical frameworks, depending on the scale. More generally, we believe that the combination of graph queries and analytics is a natural one, and one that raises interesting questions regarding languages and optimisations.

\newpage
\bibliographystyle{ACM-Reference-Format}
\bibliography{biblio}


\begin{thebibliography}{47}


\ifx \showCODEN    \undefined \def \showCODEN     #1{\unskip}     \fi
\ifx \showDOI      \undefined \def \showDOI       #1{#1}\fi
\ifx \showISBNx    \undefined \def \showISBNx     #1{\unskip}     \fi
\ifx \showISBNxiii \undefined \def \showISBNxiii  #1{\unskip}     \fi
\ifx \showISSN     \undefined \def \showISSN      #1{\unskip}     \fi
\ifx \showLCCN     \undefined \def \showLCCN      #1{\unskip}     \fi
\ifx \shownote     \undefined \def \shownote      #1{#1}          \fi
\ifx \showarticletitle \undefined \def \showarticletitle #1{#1}   \fi
\ifx \showURL      \undefined \def \showURL       {\relax}        \fi
\providecommand\bibfield[2]{#2}
\providecommand\bibinfo[2]{#2}
\providecommand\natexlab[1]{#1}
\providecommand\showeprint[2][]{arXiv:#2}

\bibitem[\protect\citeauthoryear{??}{onl}{2019}]%
        {online-appendix}
 \bibinfo{year}{2019}\natexlab{}.
\newblock \bibinfo{title}{{Online Appendix}}.
\newblock
\newblock
\newblock
\shownote{\url{https://github.com/VHDG88FKL/SPARQL-Analytics}.}


\bibitem[\protect\citeauthoryear{Abiteboul, Hull, and Vianu}{Abiteboul
  et~al\mbox{.}}{1995}]%
        {AHV95}
\bibfield{author}{\bibinfo{person}{Serge Abiteboul}, \bibinfo{person}{Richard
  Hull}, {and} \bibinfo{person}{Victor Vianu}.}
  \bibinfo{year}{1995}\natexlab{}.
\newblock \bibinfo{booktitle}{\emph{Foundations of databases}}.
  Vol.~\bibinfo{volume}{8}.
\newblock \bibinfo{publisher}{Addison-Wesley Reading}.
\newblock


\bibitem[\protect\citeauthoryear{Angles, Arenas, Barcel{\'{o}}, Boncz,
  Fletcher, Gutierrez, Lindaaker, Paradies, Plantikow, Sequeda, van Rest, and
  Voigt}{Angles et~al\mbox{.}}{2018}]%
        {AnglesABBFGLPPS18}
\bibfield{author}{\bibinfo{person}{Renzo Angles}, \bibinfo{person}{Marcelo
  Arenas}, \bibinfo{person}{Pablo Barcel{\'{o}}}, \bibinfo{person}{Peter~A.
  Boncz}, \bibinfo{person}{George H.~L. Fletcher}, \bibinfo{person}{Claudio
  Gutierrez}, \bibinfo{person}{Tobias Lindaaker}, \bibinfo{person}{Marcus
  Paradies}, \bibinfo{person}{Stefan Plantikow}, \bibinfo{person}{Juan~F.
  Sequeda}, \bibinfo{person}{Oskar van Rest}, {and} \bibinfo{person}{Hannes
  Voigt}.} \bibinfo{year}{2018}\natexlab{}.
\newblock \showarticletitle{{G-CORE: A Core for Future Graph Query Languages}}.
  In \bibinfo{booktitle}{\emph{{SIGMOD}}}. \bibinfo{pages}{1421--1432}.
\newblock


\bibitem[\protect\citeauthoryear{Angles, Arenas, Barcel{\'{o}}, Hogan, Reutter,
  and Vrgoc}{Angles et~al\mbox{.}}{2017}]%
        {AnglesABHRV17}
\bibfield{author}{\bibinfo{person}{Renzo Angles}, \bibinfo{person}{Marcelo
  Arenas}, \bibinfo{person}{Pablo Barcel{\'{o}}}, \bibinfo{person}{Aidan
  Hogan}, \bibinfo{person}{Juan~L. Reutter}, {and} \bibinfo{person}{Domagoj
  Vrgoc}.} \bibinfo{year}{2017}\natexlab{}.
\newblock \showarticletitle{{Foundations of Modern Query Languages for Graph
  Databases}}.
\newblock \bibinfo{journal}{\emph{{ACM} Comput. Surv.}} \bibinfo{volume}{50},
  \bibinfo{number}{5} (\bibinfo{year}{2017}), \bibinfo{pages}{68:1--68:40}.
\newblock


\bibitem[\protect\citeauthoryear{Aranda, Polleres, and Umbrich}{Aranda
  et~al\mbox{.}}{2014}]%
        {ArandaPU14}
\bibfield{author}{\bibinfo{person}{Carlos~Buil Aranda}, \bibinfo{person}{Axel
  Polleres}, {and} \bibinfo{person}{J{\"{u}}rgen Umbrich}.}
  \bibinfo{year}{2014}\natexlab{}.
\newblock \showarticletitle{{Strategies for Executing Federated Queries in
  SPARQL1.1}}. In \bibinfo{booktitle}{\emph{International Semantic Web
  Conference (ISWC)}}. \bibinfo{publisher}{Springer},
  \bibinfo{pages}{390--405}.
\newblock


\bibitem[\protect\citeauthoryear{Battaglia, Hamrick, Bapst, Sanchez-Gonzalez,
  Zambaldi, Malinowski, Tacchetti, Raposo, Santoro, Faulkner,
  et~al\mbox{.}}{Battaglia et~al\mbox{.}}{2018}]%
        {battaglia}
\bibfield{author}{\bibinfo{person}{Peter~W Battaglia},
  \bibinfo{person}{Jessica~B Hamrick}, \bibinfo{person}{Victor Bapst},
  \bibinfo{person}{Alvaro Sanchez-Gonzalez}, \bibinfo{person}{Vinicius
  Zambaldi}, \bibinfo{person}{Mateusz Malinowski}, \bibinfo{person}{Andrea
  Tacchetti}, \bibinfo{person}{David Raposo}, \bibinfo{person}{Adam Santoro},
  \bibinfo{person}{Ryan Faulkner}, {et~al\mbox{.}}}
  \bibinfo{year}{2018}\natexlab{}.
\newblock \showarticletitle{Relational inductive biases, deep learning, and
  graph networks}.
\newblock \bibinfo{journal}{\emph{arXiv preprint arXiv:1806.01261}}
  (\bibinfo{year}{2018}).
\newblock


\bibitem[\protect\citeauthoryear{Bonatti, Decker, Polleres, and
  Presutti}{Bonatti et~al\mbox{.}}{2018}]%
        {BonattiDPP18}
\bibfield{author}{\bibinfo{person}{Piero~Andrea Bonatti},
  \bibinfo{person}{Stefan Decker}, \bibinfo{person}{Axel Polleres}, {and}
  \bibinfo{person}{Valentina Presutti}.} \bibinfo{year}{2018}\natexlab{}.
\newblock \showarticletitle{{Knowledge Graphs: New Directions for Knowledge
  Representation on the Semantic Web}}.
\newblock \bibinfo{journal}{\emph{Dagstuhl Reports}} \bibinfo{volume}{8},
  \bibinfo{number}{9} (\bibinfo{year}{2018}), \bibinfo{pages}{29--111}.
\newblock


\bibitem[\protect\citeauthoryear{Brijder, Geerts, den Bussche, and
  Weerwag}{Brijder et~al\mbox{.}}{2018}]%
        {BrijderGBW18}
\bibfield{author}{\bibinfo{person}{Robert Brijder}, \bibinfo{person}{Floris
  Geerts}, \bibinfo{person}{Jan~Van den Bussche}, {and} \bibinfo{person}{Timmy
  Weerwag}.} \bibinfo{year}{2018}\natexlab{}.
\newblock \showarticletitle{{On the Expressive Power of Query Languages for
  Matrices}}. In \bibinfo{booktitle}{\emph{International Conference on Database
  Theory (ICDT)}}. \bibinfo{publisher}{Schloss Dagstuhl},
  \bibinfo{pages}{10:1--10:17}.
\newblock


\bibitem[\protect\citeauthoryear{Cai, F{\"u}rer, and Immerman}{Cai
  et~al\mbox{.}}{1992}]%
        {CFI92}
\bibfield{author}{\bibinfo{person}{Jin-Yi Cai}, \bibinfo{person}{Martin
  F{\"u}rer}, {and} \bibinfo{person}{Neil Immerman}.}
  \bibinfo{year}{1992}\natexlab{}.
\newblock
  \showarticletitle{\href{https://people.cs.umass.edu/~immerman/pub/opt.pdf}{An
  optimal lower bound on the number of variables for graph identification}}.
\newblock \bibinfo{journal}{\emph{Combinatorica}} \bibinfo{volume}{12},
  \bibinfo{number}{4} (\bibinfo{year}{1992}), \bibinfo{pages}{389--410}.
\newblock


\bibitem[\protect\citeauthoryear{Chang}{Chang}{2018}]%
        {AirBnBKG}
\bibfield{author}{\bibinfo{person}{Spencer Chang}.}
  \bibinfo{year}{2018}\natexlab{}.
\newblock \bibinfo{title}{{Scaling Knowledge Access and Retrieval at Airbnb}}.
\newblock \bibinfo{howpublished}{AirBnB Medium Blog}.
\newblock
\newblock
\shownote{\url{https://medium.com/airbnb-engineering/scaling-knowledge-access-and-retrieval-at-airbnb-665b6ba21e95}.}


\bibitem[\protect\citeauthoryear{Ching, Edunov, Kabiljo, Logothetis, and
  Muthukrishnan}{Ching et~al\mbox{.}}{2015}]%
        {ChingEKLM15}
\bibfield{author}{\bibinfo{person}{Avery Ching}, \bibinfo{person}{Sergey
  Edunov}, \bibinfo{person}{Maja Kabiljo}, \bibinfo{person}{Dionysios
  Logothetis}, {and} \bibinfo{person}{Sambavi Muthukrishnan}.}
  \bibinfo{year}{2015}\natexlab{}.
\newblock \showarticletitle{One Trillion Edges: Graph Processing at
  Facebook-Scale}.
\newblock \bibinfo{journal}{\emph{{PVLDB}}} \bibinfo{volume}{8},
  \bibinfo{number}{12} (\bibinfo{year}{2015}), \bibinfo{pages}{1804--1815}.
\newblock


\bibitem[\protect\citeauthoryear{Corby, Faron{-}Zucker, and Gandon}{Corby
  et~al\mbox{.}}{2017}]%
        {CorbyFG17}
\bibfield{author}{\bibinfo{person}{Olivier Corby}, \bibinfo{person}{Catherine
  Faron{-}Zucker}, {and} \bibinfo{person}{Fabien Gandon}.}
  \bibinfo{year}{2017}\natexlab{}.
\newblock \showarticletitle{{LDScript: A Linked Data Script Language}}. In
  \bibinfo{booktitle}{\emph{International Semantic Web Conference (ISWC)}}.
  \bibinfo{publisher}{Springer}, \bibinfo{pages}{208--224}.
\newblock


\bibitem[\protect\citeauthoryear{DeLorimier, Kapre, Mehta, Rizzo, Eslick,
  Rubin, Uribe, Jr., and DeHon}{DeLorimier et~al\mbox{.}}{2006}]%
        {DeLorimierKMRERUKD06}
\bibfield{author}{\bibinfo{person}{Michael DeLorimier},
  \bibinfo{person}{Nachiket Kapre}, \bibinfo{person}{Nikil Mehta},
  \bibinfo{person}{Dominic Rizzo}, \bibinfo{person}{Ian Eslick},
  \bibinfo{person}{Raphael Rubin}, \bibinfo{person}{Tom{\'{a}}s~E. Uribe},
  \bibinfo{person}{Thomas F.~Knight Jr.}, {and} \bibinfo{person}{Andr{\'{e}}
  DeHon}.} \bibinfo{year}{2006}\natexlab{}.
\newblock \showarticletitle{{GraphStep: A System Architecture for Sparse-Graph
  Algorithms}}. In \bibinfo{booktitle}{\emph{{IEEE} Symposium on
  Field-Programmable Custom Computing Machines (FCCM)}}.
  \bibinfo{publisher}{{IEEE} Computer Society}, \bibinfo{pages}{143--151}.
\newblock


\bibitem[\protect\citeauthoryear{Denny~Vrande{\v{c}}i{\'{c} and Markus
  Kr{\"{o}}tzsch}}{Denny~Vrande{\v{c}}i{\'{c} and Markus
  Kr{\"{o}}tzsch}}{[n.d.]}]%
        {VrandecicK14}
\bibfield{author}{\bibinfo{person}{journal = {Commun. {ACM}} volume = {57}
  number = {10} pages = {78--85} year =~{2014} Denny~Vrande{\v{c}}i{\'{c} and
  Markus Kr{\"{o}}tzsch}, title =~{Wikidata: a free collaborative
  knowledgebase}}.} \bibinfo{year}{[n.d.]}\natexlab{}.
\newblock  (\bibinfo{year}{[n.\,d.]}).
\newblock


\bibitem[\protect\citeauthoryear{Francis, Green, Guagliardo, Libkin, Lindaaker,
  Marsault, Plantikow, Rydberg, Selmer, and Taylor}{Francis
  et~al\mbox{.}}{2018}]%
        {FrancisGGLLMPRS18}
\bibfield{author}{\bibinfo{person}{Nadime Francis}, \bibinfo{person}{Alastair
  Green}, \bibinfo{person}{Paolo Guagliardo}, \bibinfo{person}{Leonid Libkin},
  \bibinfo{person}{Tobias Lindaaker}, \bibinfo{person}{Victor Marsault},
  \bibinfo{person}{Stefan Plantikow}, \bibinfo{person}{Mats Rydberg},
  \bibinfo{person}{Petra Selmer}, {and} \bibinfo{person}{Andr{\'{e}}s Taylor}.}
  \bibinfo{year}{2018}\natexlab{}.
\newblock \showarticletitle{{Cypher: An Evolving Query Language for Property
  Graphs}}. In \bibinfo{booktitle}{\emph{International Conference on Management
  of Data (SIGMOD)}}. \bibinfo{publisher}{{ACM}}, \bibinfo{pages}{1433--1445}.
\newblock


\bibitem[\protect\citeauthoryear{Geerts}{Geerts}{2019}]%
        {Geerts19}
\bibfield{author}{\bibinfo{person}{Floris Geerts}.}
  \bibinfo{year}{2019}\natexlab{}.
\newblock \showarticletitle{{On the Expressive Power of Linear Algebra on
  Graphs}}. In \bibinfo{booktitle}{\emph{International Conference on Database
  Theory (ICDT)}}. \bibinfo{publisher}{Schloss Dagstuhl},
  \bibinfo{pages}{7:1--7:19}.
\newblock


\bibitem[\protect\citeauthoryear{Gonzalez, Low, Gu, Bickson, and
  Guestrin}{Gonzalez et~al\mbox{.}}{2012}]%
        {GonzalezLGBG12}
\bibfield{author}{\bibinfo{person}{Joseph~E. Gonzalez},
  \bibinfo{person}{Yucheng Low}, \bibinfo{person}{Haijie Gu},
  \bibinfo{person}{Danny Bickson}, {and} \bibinfo{person}{Carlos Guestrin}.}
  \bibinfo{year}{2012}\natexlab{}.
\newblock \showarticletitle{{PowerGraph: Distributed Graph-Parallel Computation
  on Natural Graphs}}. In \bibinfo{booktitle}{\emph{10th {USENIX} Symposium on
  Operating Systems Design and Implementation, {OSDI} 2012, Hollywood, CA, USA,
  October 8-10, 2012}}. \bibinfo{publisher}{{USENIX} Association},
  \bibinfo{pages}{17--30}.
\newblock


\bibitem[\protect\citeauthoryear{Harris, Seaborne, and Prud'hommeaux}{Harris
  et~al\mbox{.}}{2013}]%
        {sparql11}
\bibfield{author}{\bibinfo{person}{Steve Harris}, \bibinfo{person}{Andy
  Seaborne}, {and} \bibinfo{person}{Eric Prud'hommeaux}.}
  \bibinfo{year}{2013}\natexlab{}.
\newblock \bibinfo{title}{{SPARQL 1.1 Query Language}}.
\newblock \bibinfo{howpublished}{W3C Recommendation}.
\newblock
\newblock
\shownote{\url{https://www.w3.org/TR/sparql11-query/}.}


\bibitem[\protect\citeauthoryear{He, Chen, and Agarwal}{He
  et~al\mbox{.}}{2016}]%
        {LinkedInKG}
\bibfield{author}{\bibinfo{person}{Qi He}, \bibinfo{person}{Bee-Chung Chen},
  {and} \bibinfo{person}{Deepak Agarwal}.} \bibinfo{year}{2016}\natexlab{}.
\newblock \bibinfo{title}{{Building The LinkedIn Knowledge Graph}}.
\newblock \bibinfo{howpublished}{LinkedIn Blog}.
\newblock
\newblock
\shownote{\url{https://engineering.linkedin.com/blog/2016/10/building-the-linkedin-knowledge-graph}.}


\bibitem[\protect\citeauthoryear{Hutchison, Howe, and Suciu}{Hutchison
  et~al\mbox{.}}{2017}]%
        {HutchisonHS17}
\bibfield{author}{\bibinfo{person}{Dylan Hutchison}, \bibinfo{person}{Bill
  Howe}, {and} \bibinfo{person}{Dan Suciu}.} \bibinfo{year}{2017}\natexlab{}.
\newblock \showarticletitle{{LaraDB: A Minimalist Kernel for Linear and
  Relational Algebra Computation}}. In \bibinfo{booktitle}{\emph{{ACM} {SIGMOD}
  Workshop on Algorithms and Systems for MapReduce and Beyond
  (BeyondMR@SIGMOD)}}. \bibinfo{publisher}{{ACM}}, \bibinfo{pages}{2:1--2:10}.
\newblock


\bibitem[\protect\citeauthoryear{Kostylev, Reutter, Romero, and Vrgoc}{Kostylev
  et~al\mbox{.}}{2015}]%
        {KostylevR0V15}
\bibfield{author}{\bibinfo{person}{Egor~V. Kostylev}, \bibinfo{person}{Juan~L.
  Reutter}, \bibinfo{person}{Miguel Romero}, {and} \bibinfo{person}{Domagoj
  Vrgoc}.} \bibinfo{year}{2015}\natexlab{}.
\newblock \showarticletitle{{SPARQL with Property Paths}}. In
  \bibinfo{booktitle}{\emph{International Semantic Web Conference (ISWC)}}.
  \bibinfo{publisher}{Springer}, \bibinfo{pages}{3--18}.
\newblock


\bibitem[\protect\citeauthoryear{Krepska, Kielmann, Fokkink, and Bal}{Krepska
  et~al\mbox{.}}{2011}]%
        {KrepskaKFB11}
\bibfield{author}{\bibinfo{person}{Elzbieta Krepska}, \bibinfo{person}{Thilo
  Kielmann}, \bibinfo{person}{Wan Fokkink}, {and} \bibinfo{person}{Henri~E.
  Bal}.} \bibinfo{year}{2011}\natexlab{}.
\newblock \showarticletitle{{HipG}: parallel processing of large-scale graphs}.
\newblock \bibinfo{journal}{\emph{Operating Systems Review}}
  \bibinfo{volume}{45}, \bibinfo{number}{2} (\bibinfo{year}{2011}),
  \bibinfo{pages}{3--13}.
\newblock


\bibitem[\protect\citeauthoryear{Krishnan}{Krishnan}{2018}]%
        {AmazonKG}
\bibfield{author}{\bibinfo{person}{Arun Krishnan}.}
  \bibinfo{year}{2018}\natexlab{}.
\newblock \bibinfo{title}{{Making search easier: How Amazon's Product Graph is
  helping customers find products more easily}}.
\newblock \bibinfo{howpublished}{Amazon Blog}.
\newblock
\newblock
\shownote{\url{https://blog.aboutamazon.com/innovation/making-search-easier}.}


\bibitem[\protect\citeauthoryear{LDBC}{LDBC}{2019}]%
        {graphalytics}
\bibfield{author}{\bibinfo{person}{LDBC}.} \bibinfo{year}{2019}\natexlab{}.
\newblock \bibinfo{title}{{Graphalytics Benchmark Suite}}.
\newblock
\newblock
\newblock
\shownote{\url{https://graphalytics.org/}.}


\bibitem[\protect\citeauthoryear{Low, Gonzalez, Kyrola, Bickson, Guestrin, and
  Hellerstein}{Low et~al\mbox{.}}{2014}]%
        {LowGKBGH14}
\bibfield{author}{\bibinfo{person}{Yucheng Low}, \bibinfo{person}{Joseph~E.
  Gonzalez}, \bibinfo{person}{Aapo Kyrola}, \bibinfo{person}{Danny Bickson},
  \bibinfo{person}{Carlos Guestrin}, {and} \bibinfo{person}{Joseph~M.
  Hellerstein}.} \bibinfo{year}{2014}\natexlab{}.
\newblock \showarticletitle{GraphLab: {A} New Framework For Parallel Machine
  Learning}.
\newblock \bibinfo{journal}{\emph{CoRR}}  \bibinfo{volume}{abs/1408.2041}
  (\bibinfo{year}{2014}).
\newblock
\urldef\tempurl%
\url{http://arxiv.org/abs/1408.2041}
\showURL{%
\tempurl}


\bibitem[\protect\citeauthoryear{Malewicz, Austern, Bik, Dehnert, Horn, Leiser,
  and Czajkowski}{Malewicz et~al\mbox{.}}{2010}]%
        {MalewiczABDHLC10}
\bibfield{author}{\bibinfo{person}{Grzegorz Malewicz},
  \bibinfo{person}{Matthew~H. Austern}, \bibinfo{person}{Aart J.~C. Bik},
  \bibinfo{person}{James~C. Dehnert}, \bibinfo{person}{Ilan Horn},
  \bibinfo{person}{Naty Leiser}, {and} \bibinfo{person}{Grzegorz Czajkowski}.}
  \bibinfo{year}{2010}\natexlab{}.
\newblock \showarticletitle{{Pregel: a system for large-scale graph
  processing}}. In \bibinfo{booktitle}{\emph{Proceedings of the {ACM} {SIGMOD}
  International Conference on Management of Data, {SIGMOD} 2010, Indianapolis,
  Indiana, USA, June 6-10, 2010}}. \bibinfo{publisher}{{ACM} Press},
  \bibinfo{pages}{135--146}.
\newblock


\bibitem[\protect\citeauthoryear{Malyshev, Kr{\"{o}}tzsch, Gonz{\'{a}}lez,
  Gonsior, and Bielefeldt}{Malyshev et~al\mbox{.}}{2018}]%
        {MalyshevKGGB18}
\bibfield{author}{\bibinfo{person}{Stanislav Malyshev}, \bibinfo{person}{Markus
  Kr{\"{o}}tzsch}, \bibinfo{person}{Larry Gonz{\'{a}}lez},
  \bibinfo{person}{Julius Gonsior}, {and} \bibinfo{person}{Adrian Bielefeldt}.}
  \bibinfo{year}{2018}\natexlab{}.
\newblock \showarticletitle{{Getting the Most Out of Wikidata: Semantic
  Technology Usage in Wikipedia's Knowledge Graph}}. In
  \bibinfo{booktitle}{\emph{International Semantic Web Conference (ISWC)}}.
  \bibinfo{publisher}{Springer}, \bibinfo{pages}{376--394}.
\newblock


\bibitem[\protect\citeauthoryear{Miller}{Miller}{2013}]%
        {Miller13}
\bibfield{author}{\bibinfo{person}{Justin~J. Miller}.}
  \bibinfo{year}{2013}\natexlab{}.
\newblock \showarticletitle{{Graph Database Applications and Concepts with
  Neo4j}}. In \bibinfo{booktitle}{\emph{Southern Association for Information
  Systems Conference (SAIS)}}. \bibinfo{publisher}{AIS eLibrary}.
\newblock


\bibitem[\protect\citeauthoryear{Morris, Ritzert, Fey, Hamilton, Lenssen,
  Rattan, and Grohe}{Morris et~al\mbox{.}}{2018}]%
        {morris}
\bibfield{author}{\bibinfo{person}{Christopher Morris}, \bibinfo{person}{Martin
  Ritzert}, \bibinfo{person}{Matthias Fey}, \bibinfo{person}{William~L
  Hamilton}, \bibinfo{person}{Jan~Eric Lenssen}, \bibinfo{person}{Gaurav
  Rattan}, {and} \bibinfo{person}{Martin Grohe}.}
  \bibinfo{year}{2018}\natexlab{}.
\newblock \showarticletitle{Weisfeiler and leman go neural: Higher-order graph
  neural networks}.
\newblock \bibinfo{journal}{\emph{arXiv preprint arXiv:1810.02244}}
  (\bibinfo{year}{2018}).
\newblock


\bibitem[\protect\citeauthoryear{Page, Brin, Motwani, and Winograd}{Page
  et~al\mbox{.}}{1999}]%
        {pagerank}
\bibfield{author}{\bibinfo{person}{Lawrence Page}, \bibinfo{person}{Sergey
  Brin}, \bibinfo{person}{Rajeev Motwani}, {and} \bibinfo{person}{Terry
  Winograd}.} \bibinfo{year}{1999}\natexlab{}.
\newblock \bibinfo{booktitle}{\emph{{The PageRank citation ranking: Bringing
  order to the Web}}}.
\newblock \bibinfo{type}{{T}echnical {R}eport}. \bibinfo{institution}{Stanford
  InfoLab}.
\newblock


\bibitem[\protect\citeauthoryear{P{\'e}rez, Arenas, and Gutierrez}{P{\'e}rez
  et~al\mbox{.}}{2009}]%
        {perez2009semantics}
\bibfield{author}{\bibinfo{person}{Jorge P{\'e}rez}, \bibinfo{person}{Marcelo
  Arenas}, {and} \bibinfo{person}{Claudio Gutierrez}.}
  \bibinfo{year}{2009}\natexlab{}.
\newblock \showarticletitle{Semantics and complexity of SPARQL}.
\newblock \bibinfo{journal}{\emph{ACM Transactions on Database Systems (TODS)}}
  \bibinfo{volume}{34}, \bibinfo{number}{3} (\bibinfo{year}{2009}),
  \bibinfo{pages}{16}.
\newblock


\bibitem[\protect\citeauthoryear{Pittman, Srivastava, Hewavitharana, Kale, and
  Mansour}{Pittman et~al\mbox{.}}{2017}]%
        {eBayKG}
\bibfield{author}{\bibinfo{person}{RJ Pittman}, \bibinfo{person}{Amit
  Srivastava}, \bibinfo{person}{Sanjika Hewavitharana},
  \bibinfo{person}{Ajinkya Kale}, {and} \bibinfo{person}{Saab Mansour}.}
  \bibinfo{year}{2017}\natexlab{}.
\newblock \bibinfo{title}{{Cracking the Code on Conversational Commerce}}.
\newblock \bibinfo{howpublished}{eBay Blog}.
\newblock
\newblock
\shownote{\url{https://www.ebayinc.com/stories/news/cracking-the-code-on-conversational-commerce/}.}


\bibitem[\protect\citeauthoryear{Reutter, Soto, and Vrgoc}{Reutter
  et~al\mbox{.}}{2015}]%
        {ReutterSV15}
\bibfield{author}{\bibinfo{person}{Juan~L. Reutter},
  \bibinfo{person}{Adri{\'{a}}n Soto}, {and} \bibinfo{person}{Domagoj Vrgoc}.}
  \bibinfo{year}{2015}\natexlab{}.
\newblock \showarticletitle{Recursion in {SPARQL}}. In
  \bibinfo{booktitle}{\emph{International Semantic Web Conference (ISWC)}}.
  \bibinfo{publisher}{Springer}, \bibinfo{pages}{19--35}.
\newblock


\bibitem[\protect\citeauthoryear{Rodriguez}{Rodriguez}{2015}]%
        {Rodriguez15}
\bibfield{author}{\bibinfo{person}{Marko~A. Rodriguez}.}
  \bibinfo{year}{2015}\natexlab{}.
\newblock \showarticletitle{{The Gremlin graph traversal machine and
  language}}. In \bibinfo{booktitle}{\emph{Symposium on Database Programming
  Languages (DBPL)}}. \bibinfo{publisher}{{ACM}}, \bibinfo{pages}{1--10}.
\newblock


\bibitem[\protect\citeauthoryear{Scarselli, Gori, Tsoi, Hagenbuchner, and
  Monfardini}{Scarselli et~al\mbox{.}}{2009}]%
        {ScarselliGTHM09}
\bibfield{author}{\bibinfo{person}{Franco Scarselli}, \bibinfo{person}{Marco
  Gori}, \bibinfo{person}{Ah~Chung Tsoi}, \bibinfo{person}{Markus
  Hagenbuchner}, {and} \bibinfo{person}{Gabriele Monfardini}.}
  \bibinfo{year}{2009}\natexlab{}.
\newblock \showarticletitle{{The Graph Neural Network Model}}.
\newblock \bibinfo{journal}{\emph{{IEEE} Trans. Neural Networks}}
  \bibinfo{volume}{20}, \bibinfo{number}{1} (\bibinfo{year}{2009}),
  \bibinfo{pages}{61--80}.
\newblock


\bibitem[\protect\citeauthoryear{Senanayake, Piraveenan, and Zomaya}{Senanayake
  et~al\mbox{.}}{2015}]%
        {Senanayake15}
\bibfield{author}{\bibinfo{person}{Upul Senanayake}, \bibinfo{person}{Mahendra
  Piraveenan}, {and} \bibinfo{person}{Albert Zomaya}.}
  \bibinfo{year}{2015}\natexlab{}.
\newblock \showarticletitle{{The Pagerank-Index: Going beyond Citation Counts
  in Quantifying Scientific Impact of Researchers}}.
\newblock \bibinfo{journal}{\emph{PLOS ONE}} \bibinfo{volume}{10},
  \bibinfo{number}{8} (\bibinfo{date}{08} \bibinfo{year}{2015}),
  \bibinfo{pages}{1--34}.
\newblock


\bibitem[\protect\citeauthoryear{Shao, Wang, and Li}{Shao
  et~al\mbox{.}}{2013}]%
        {ShaoWL13}
\bibfield{author}{\bibinfo{person}{Bin Shao}, \bibinfo{person}{Haixun Wang},
  {and} \bibinfo{person}{Yatao Li}.} \bibinfo{year}{2013}\natexlab{}.
\newblock \showarticletitle{Trinity: a distributed graph engine on a memory
  cloud}. In \bibinfo{booktitle}{\emph{{SIGMOD} International Conference on
  Management of Data (SIGMOD)}}. \bibinfo{publisher}{{ACM}},
  \bibinfo{pages}{505--516}.
\newblock


\bibitem[\protect\citeauthoryear{Shrivastava}{Shrivastava}{2017}]%
        {BingKG}
\bibfield{author}{\bibinfo{person}{Saurabh Shrivastava}.}
  \bibinfo{year}{2017}\natexlab{}.
\newblock \bibinfo{title}{{Bring rich knowledge of people, places, things and
  local businesses to your apps}}.
\newblock \bibinfo{howpublished}{Bing Blogs}.
\newblock
\newblock
\shownote{\url{https://blogs.bing.com/search-quality-insights/2017-07/bring-rich-knowledge-of-people-places-things-and-local-businesses-to-your-apps}.}


\bibitem[\protect\citeauthoryear{Singhal}{Singhal}{2012}]%
        {GoogleKG}
\bibfield{author}{\bibinfo{person}{Amit Singhal}.}
  \bibinfo{year}{2012}\natexlab{}.
\newblock \bibinfo{title}{{Introducing the Knowledge Graph: things, not
  strings}}.
\newblock \bibinfo{howpublished}{Google Blog}.
\newblock
\newblock
\shownote{\url{https://www.blog.google/products/search/introducing-knowledge-graph-things-not/}.}


\bibitem[\protect\citeauthoryear{Stutz, Strebel, and Bernstein}{Stutz
  et~al\mbox{.}}{2016}]%
        {signalcollect}
\bibfield{author}{\bibinfo{person}{Philip Stutz}, \bibinfo{person}{Daniel
  Strebel}, {and} \bibinfo{person}{Abraham Bernstein}.}
  \bibinfo{year}{2016}\natexlab{}.
\newblock \showarticletitle{{Signal/Collect12}}.
\newblock \bibinfo{journal}{\emph{Semantic Web Journal}} \bibinfo{volume}{7},
  \bibinfo{number}{2} (\bibinfo{year}{2016}), \bibinfo{pages}{139--166}.
\newblock


\bibitem[\protect\citeauthoryear{Urzua and Guti{\'{e}}rrez}{Urzua and
  Guti{\'{e}}rrez}{2019}]%
        {Urzua019}
\bibfield{author}{\bibinfo{person}{Valentina Urzua} {and}
  \bibinfo{person}{Claudio Guti{\'{e}}rrez}.} \bibinfo{year}{2019}\natexlab{}.
\newblock \showarticletitle{{Linear Recursion in G-CORE}}. In
  \bibinfo{booktitle}{\emph{Alberto Mendelzon International Workshop on
  Foundations of Data Management (AMW)}}, Vol.~\bibinfo{volume}{2369}.
  \bibinfo{publisher}{CEUR-WS.org}.
\newblock


\bibitem[\protect\citeauthoryear{Wu, Pan, Chen, Long, Zhang, and Yu}{Wu
  et~al\mbox{.}}{2019}]%
        {abs-1901-00596}
\bibfield{author}{\bibinfo{person}{Zonghan Wu}, \bibinfo{person}{Shirui Pan},
  \bibinfo{person}{Fengwen Chen}, \bibinfo{person}{Guodong Long},
  \bibinfo{person}{Chengqi Zhang}, {and} \bibinfo{person}{Philip~S. Yu}.}
  \bibinfo{year}{2019}\natexlab{}.
\newblock \showarticletitle{{A Comprehensive Survey on Graph Neural Networks}}.
\newblock \bibinfo{journal}{\emph{CoRR}}  \bibinfo{volume}{abs/1901.00596}
  (\bibinfo{year}{2019}).
\newblock


\bibitem[\protect\citeauthoryear{Xin, Gonzalez, Franklin, and Stoica}{Xin
  et~al\mbox{.}}{2013a}]%
        {XinGFS13}
\bibfield{author}{\bibinfo{person}{Reynold~S. Xin}, \bibinfo{person}{Joseph~E.
  Gonzalez}, \bibinfo{person}{Michael~J. Franklin}, {and} \bibinfo{person}{Ion
  Stoica}.} \bibinfo{year}{2013}\natexlab{a}.
\newblock \showarticletitle{{GraphX: a resilient distributed graph system on
  spark}}. In \bibinfo{booktitle}{\emph{First International Workshop on Graph
  Data Management Experiences and Systems, {GRADES} 2013, co-loated with
  {SIGMOD/PODS} 2013, New York, NY, USA, June 24, 2013}}.
  \bibinfo{publisher}{{ACM} Press}.
\newblock


\bibitem[\protect\citeauthoryear{Xin, Rosen, Zaharia, Franklin, Shenker, and
  Stoica}{Xin et~al\mbox{.}}{2013b}]%
        {XinRZFSS13}
\bibfield{author}{\bibinfo{person}{Reynold~S. Xin}, \bibinfo{person}{Josh
  Rosen}, \bibinfo{person}{Matei Zaharia}, \bibinfo{person}{Michael~J.
  Franklin}, \bibinfo{person}{Scott Shenker}, {and} \bibinfo{person}{Ion
  Stoica}.} \bibinfo{year}{2013}\natexlab{b}.
\newblock \showarticletitle{{Shark: {SQL} and rich analytics at scale}}. In
  \bibinfo{booktitle}{\emph{Proceedings of the {ACM} {SIGMOD} International
  Conference on Management of Data, {SIGMOD} 2013, New York, NY, USA, June
  22-27, 2013}}. \bibinfo{publisher}{{ACM} Press}, \bibinfo{pages}{13--24}.
\newblock


\bibitem[\protect\citeauthoryear{Xu, Hu, Leskovec, and Jegelka}{Xu
  et~al\mbox{.}}{2019}]%
        {XuHLJ19}
\bibfield{author}{\bibinfo{person}{Keyulu Xu}, \bibinfo{person}{Weihua Hu},
  \bibinfo{person}{Jure Leskovec}, {and} \bibinfo{person}{Stefanie Jegelka}.}
  \bibinfo{year}{2019}\natexlab{}.
\newblock \showarticletitle{{How Powerful are Graph Neural Networks?}}. In
  \bibinfo{booktitle}{\emph{International Conference on Learning
  Representations ({ICLR})}}. \bibinfo{publisher}{OpenReview.net}.
\newblock


\bibitem[\protect\citeauthoryear{Zaharia, Xin, Wendell, Das, Armbrust, Dave,
  Meng, Rosen, Venkataraman, Franklin, Ghodsi, Gonzalez, Shenker, and
  Stoica}{Zaharia et~al\mbox{.}}{2016}]%
        {spark}
\bibfield{author}{\bibinfo{person}{Matei Zaharia}, \bibinfo{person}{Reynold~S.
  Xin}, \bibinfo{person}{Patrick Wendell}, \bibinfo{person}{Tathagata Das},
  \bibinfo{person}{Michael Armbrust}, \bibinfo{person}{Ankur Dave},
  \bibinfo{person}{Xiangrui Meng}, \bibinfo{person}{Josh Rosen},
  \bibinfo{person}{Shivaram Venkataraman}, \bibinfo{person}{Michael~J.
  Franklin}, \bibinfo{person}{Ali Ghodsi}, \bibinfo{person}{Joseph Gonzalez},
  \bibinfo{person}{Scott Shenker}, {and} \bibinfo{person}{Ion Stoica}.}
  \bibinfo{year}{2016}\natexlab{}.
\newblock \showarticletitle{{Apache Spark: a unified engine for big data
  processing}}.
\newblock \bibinfo{journal}{\emph{Commun. {ACM}}} \bibinfo{volume}{59},
  \bibinfo{number}{11} (\bibinfo{year}{2016}), \bibinfo{pages}{56--65}.
\newblock


\bibitem[\protect\citeauthoryear{Zeng, Yang, Wang, Shao, and Wang}{Zeng
  et~al\mbox{.}}{2013}]%
        {ZengYWSW13}
\bibfield{author}{\bibinfo{person}{Kai Zeng}, \bibinfo{person}{Jiacheng Yang},
  \bibinfo{person}{Haixun Wang}, \bibinfo{person}{Bin Shao}, {and}
  \bibinfo{person}{Zhongyuan Wang}.} \bibinfo{year}{2013}\natexlab{}.
\newblock \showarticletitle{{A Distributed Graph Engine for Web Scale RDF
  Data}}.
\newblock \bibinfo{journal}{\emph{{PVLDB}}} \bibinfo{volume}{6},
  \bibinfo{number}{4} (\bibinfo{year}{2013}), \bibinfo{pages}{265--276}.
\newblock


\end{thebibliography}


\appendix
\onecolumn
\section{Appendix: Proofs}

\subsection{Proof of Theorem \ref{teo:tc}}

Let $M = (Q,F,\Sigma \cup \{B\},q_0,\delta$ be a deterministic Turing machine, where $Q = \{q_o,\dots,q_m\}$ is the set of states, there is a single final state $F = \{q_m\}$, 
$\Sigma$ is the alphabet, $B$ is the blank node covering all cells and $\delta$ is the transition function. Without loss of generality, and for readability, 
we assume that $\Sigma = \{0,1\}$ and that $\delta$ does not define transitions for $q_m$. 
Let also $w = a_0,\dots,a_n$ be a binary string. We construct a graph $G$ and a SPARQAL procedure $P$ such that $M$ accepts $w$ if and only if $P$ returns a non-empty mapping. 

Let us first assume that all states in $Q$ and characters $0$, $1$, $B$ are represented by IRIs, and that we use IRIs \text{:right} and \text{:left}.
Define $T_\delta$ as a set of tuples of arity $5$ containing one tuple $(q,a,q',b,d)$ for each 
transition in $\delta$ of the form $\delta(q,a) = (q',b,d)$, for $d \in \text{:right},\text{:left}$.

For readability we will not make the distinction between graph and program, and rather initialize everything in the program. But the construction can be easily adapted 
so that the input is not coded directly in the program but is queried from a graph. 
The procedure $P$ consists of the following groups of statements. 

\paralist{Initialization}

First group of statements are in charge of initialising some of the solution variables. The idea of variable \texttt{transition} is to 
store the transitions of $M$. Solution variable \texttt{current} stores the content of the current cell that $M$ is pointing on, and the 
current state of the run. Solution variables \texttt{positive\_cells} and \texttt{negative\_cells} store, respectively, all cells to the right of 
the head of $M$ and all cells to the left of the head of $M$. Of course, the tape is infinite, but we only need to store cells we have already visited. 

\begin{lstlisting}
LET transition = ( 
  SELECT ?oldstate ?oldsymbol ?newstate ?newsymbol ?direction WHERE { 
    VALUES (?oldstate ?oldsymbol ?newstate ?newsymbol ?direction) {$T_\delta$} 
  } 
);
\end{lstlisting}

\begin{lstlisting}
LET current = ( 
  SELECT  ?c_symbol ?c_state WHERE { 
    VALUES ( ?c_symbol ?c_state) {(a0, q0} 
  } 
);
\end{lstlisting}

\begin{lstlisting}
LET positive_cells = ( 
  SELECT ?p_pos ?p_symbol WHERE { 
    VALUES (?p_pos ?p_symbol ) {(1, a1),...,(n,an)} 
  } 
);
\end{lstlisting}

\paralist{Loop} The loop phase of the procedure is as follows: 

\begin{lstlisting}
DO (
  S1
  S2
  S3
  S4
) WHILE ( C );
\end{lstlisting}

\noindent
Where all inner statements and conditions are defined next. The idea is that queries are used to check when the transition demands moving to the right or to the left, and depending 
on these values we update the cells accordingly. We use \texttt{new\_current} as a temporal variable that will store the pointed cell and state of the machine in the next step of the run. 

\paralist{Statement \texttt{S1}}

\begin{lstlisting}
LET new_current = (
  SELECT ?c_symbol ?c_state WHERE {
    SELECT (?newstate AS ?c_state) WHERE { 
      QVALUES(transition)
      QVALUES(current)
      FILTER(?oldstate=?c_state && ?oldsymbol=?c_symbol) 
    } . 
    SELECT (?symbol AS ?c_symbol) WHERE {
      QVALUES(positive_cells)
      FILTER(?p_pos = 1)
      BIND(IF(!bound(?p_pos),"B",?p_symbol) AS ?symbol)
    } 
  } 
);
\end{lstlisting}

\paralist{Statement \texttt{S2}}

\begin{lstlisting}
LET positive_cells = (
  SELECT ?p_pos ?p_symbol WHERE {
    {
      SELECT (?p_pos -1 AS ?p_pos) ?p_symbol WHERE {
        QVALUES(positive_cells)
        QVALUES(transition)
	    QVALUES(current) 
        FILTER(?oldstate=?c_state && ?oldsymbol=?c_symbol) 
        FILTER(?direction=:right) 
	    FILTER(?p_pos>1) 
      }
    } UNION
    { 
      SELECT (?p_pos + 1 AS ?p_pos) ?p_symbol WHERE {
        QVALUES(positive_cells) 
        QVALUES(transition)
        QVALUES(current) 
        FILTER(?oldstate=?c_state && ?oldsymbol=?c_symbol) 
        FILTER(?direction=:left) 
      }
    } UNION
    {
      SELECT (1 AS ?p_pos) (?newsymbol as ?p_symbol) WHERE {
        QVALUES(transition)
        QVALUES(current) 
        FILTER(?oldstate=?c_state && ?oldsymbol=?c_symbol) 
        FILTER(?direction=:left) 
      }
    }
  }
);
\end{lstlisting}

\paralist{Statement \texttt{S3}}

\begin{lstlisting}
LET negative_cells = (
  SELECT ?n_pos ?n_symbol WHERE {
    {
      SELECT (?n_pos + 1 AS ?n_pos) ?n_symbol WHERE {
        QVALUES(negative_cells) 
        QVALUES(transition)
        QVALUES(current) 
        FILTER(?oldstate=?c_state && ?oldsymbol=?c_symbol) 
        FILTER(?direction=:left) 
        FILTER(?n_pos<-1) 
      }
    } UNION
    {
      SELECT (?n_pos - 1 AS ?n_pos) ?n_symbol WHERE {
        QVALUES(negative_cells) 
        QVALUES(transition)
        QVALUES(current) 
        FILTER(?oldstate=?c_state && ?oldsymbol=?c_symbol) 
        FILTER(?direction =:right)
      } 
    } UNION
    {
      SELECT (-1 AS ?n_pos) (?newsymbol AS ?n_symbol) WHERE {
        QVALUES(transition)
        QVALUES(current) 
        FILTER(?oldstate=?c_state && ?oldsymbol=?c_symbol) 
        FILTER(?direction =:right) 
      }
    }
  }
);
\end{lstlisting}

\paralist{Statement \texttt{S4}}

\begin{lstlisting}
LET current = (
  SELECT ?c_pos ?c_symbol ?c_state WHERE { QVALUES(new_current) }
);
\end{lstlisting}

\paralist{Condition \texttt{C}}
\begin{lstlisting}
ASK {
  QVALUES(transition)
  QVALUES(current) 
  FILTER(?oldstate=?c_stat && ?oldsymbol=?c_symbol) 
}
\end{lstlisting}

\paralist{Return} Finally, below the loop, we return the state.

\begin{lstlisting}
LET state = (
  SELECT ?state WHERE { QVALUES(current) FILTER(?c_state = :qm) }
);
RETURN(state);
\end{lstlisting}

One can check that this program effectively returns a non-empty mapping if and only if the procedure $P$ terminates and variable \texttt{current} stores the state $q_m$. 
In turn, this happens if and only if $M$ accepts on the input. This finishes the proof.

\subsection{Proof of Proposition \ref{prop-complexity}}

We have already discussed how SPARQAL programs can be evaluated in PSPACE when they do not invent new values: all we need to store is 
(1) the current state of all variables, (2) the previous state of variables in fixed-point clauses, and (3) the current number of iterations for the case of loops with a max number (which is bounded by the query, as we do not need more iterations that  the number stated. Additionally, SPARQL queries can themselves be computed in PSPACE, which gives us the upper bound. 

For the lower bound we can use the construction in Theorem \ref{teo:tc}. Because we now that the machine $M$ runs in PSPACE, the number of cells visited is bounded by 
a number which depends on the elements on the graph. Let then $|G|$ be the size of the graph, and assume that $n = |G|^k$ is the number of maximum cells visited in any computation of $M$ over a graph with size $|G|$. 
The first thing we need is to construct a linear order from the elements of the graph, which we will store in a solution variable \texttt{order}. We can do this with a do-while iteration that keeps adding elements until there are no more to add. 
We can then extend this linear order into an order of 2k tuples, which will be stored in a solution variable \texttt{full-order}. With this full order we can now pre-compute all 
possible $n$ cells that may be visited by $M$ in solution variables \texttt{positive\_cells} and \texttt{negative\_cells}. We cannot use a numeric position anymore, but we 
can use our tuples in full order as the position. 
With these cells precomputed, we need to invoke the rest of the procedure. However, the last modification we make is that all arithmetic is replaced by the appropriate 
operation that uses our linear order.

\subsection{Proof of Theorem \ref{theo-gnns}}

The first item is shown by induction. On the first step, the labels of the WL test and the ones stored in variable \texttt{vector} coincide, and thus the first bullet is clearly satisfied. 
Now assume that on the $i$-th iteration of the program, the same label in \texttt{vector} is assigned to nodes with the same label in the WL test. 
Going from iteration $i$ to iteration $i+1$, if there are nodes in which $a$ and $b$ have the same label, it must be because (i) they had the same label in iteration 
$i$, (ii) their neighbours define an isomorphism, and (iii) their neighbours had the same label as well. Now if a pair $a$ and $b$ of nodes have different label in \texttt{vector}, it must be because 
queries $Q_a$ and $Q_b$ computed by Map draw different values. But this contradicts the fact that their neighbours are isomorphic and each of them have the same label. 

For the second bullet, all we need is to find an injective function so that a neighbourhood is mapped to this value. We can do this using group concatenation in SPARQL 
as follows (for readability we assume that 
the graph has just one type of property $:p$, apart from the label, but this can of course be extended). 

\begin{lstlisting}
LET var_1 = ( SELECT ?v ?neighbour WHERE {?v :p ?neighbour}); 
LET vector = (
  SELECT ?v ?lab WHERE { ?v rdfs:label ?lab }); 
DO ( 
 LET vector = (
     SELECT (?node AS ?v) (GROUP_CONCAT(?n_lab; SEPARATOR=", ") AS ?lab)
     WHERE {SELECT (?v AS ?node) (?neighbour AS ?v) WHERE {QVALUES(var\_1)} }. 
         {QVALUES(vector)} . 
         FILTER (?neighbour = ?v)
 );
) WHILE ( condition );
RETURN(vector);
\end{lstlisting}

\end{document}